\documentclass[12pt]{iopart}
\usepackage{graphicx}
\usepackage{xcolor}
\usepackage{float} %To force images at specific places
\graphicspath{{Images1/}}

\begin{document}
\title[Quantum bridges in phase space]{Quantum bridges in phase space: Interference and nonclassicality in strong-field enhanced % molecular
ionisation}
\author{
  H. Chomet, D. Sarkar and  C. Figueira de Morisson Faria
  }
\address{Department of Physics and Astronomy, University College London, Gower street, London WC1E 6BT, UK}
\date{\today}
\begin{abstract}
We perform a phase-space analysis of strong-field enhanced ionisation in molecules, with emphasis on quantum-interference effects. Using Wigner quasi-probability distributions and the quantum Liouville equation, we show that the momentum gates reported in a previous publication [N. Takemoto and A. Becker, Phys. Rev. A \textbf{84}, 023401 (2011)] may occur for static driving fields, and even for no external field at all. Their primary cause is an interference-induced bridging mechanism that occurs if both wells in the molecule are populated.  In the phase-space regions for which quantum bridges occur, the Wigner functions perform a clockwise rotation whose period is intrinsic to the molecule. This evolution is essentially non-classical and non-adiabatic, as it does not follow equienergy curves or field gradients. Quasi-probability transfer via quantum bridges is favoured if the electron's initial state is either spatially delocalised, or situated at the upfield molecular well. Enhanced ionisation results from the interplay of this cyclic motion, adiabatic tunnel ionisation and population trapping. Optimal conditions require minimising population trapping and using the bridging mechanism to feed into ionisation pathways along the field gradient.
\end{abstract}
\maketitle

\section{Introduction}
\label{sec:intro}
Strong-field ionisation is one of the most studied phenomena that occur when matter interacts with intense laser fields. 
Not only does it invite fundamental questions, such as the measurement and the criteria for determining tunnelling times \cite{Eckle2008,Shafir2012,Landsman2015,Liu2017,Camus2017,Ni2018}, but, in addition, it gave rise to whole research areas. Examples are laser-induced electron diffraction \cite{Zuo1993}, or ultrafast photoelectron holography \cite{Spanner2004,HuismansScience2011}; for a review see \cite{Faria2019}. Because of the vast number of applications to attosecond imaging of matter, strong-field ionisation in molecules has become hugely popular since the mid 2000s. 

A peculiar and well-known effect that occurs for stretched molecules in strong, low-frequency fields is enhanced ionisation. It consists of a sharp increase  in the ionisation rate around specific inter-nuclear separations, typically larger than the equilibrium distance. Since its first prediction \cite{Zuo1995}, enhanced ionisation has been calculated and measured in myriad systems. These include diatomic \cite{Codling1989,Chelkowski1995} molecular species such as $\mathrm{H}_2$ \cite{Seideman1995,Bandrauk1999,Pav2005,Suzuki2007,Takemo2011,Takemoto2010}, $\mathrm{I}_2$ \cite{Strickland1992,Constant1996,Chen2011,Chen2012}, and $\mathrm{Cl}_2$ \cite{Schmidt1994}, tri- \cite{Suzuki2007,Bocharova2011,Paul2019} and polyatomic molecules \cite{DeWitt1998,Erattupuzha2017}. It has also been used as a mean to highlight electron nuclear coupling of degrees of freedom in photoelectron holography \cite{Mi2017}.
 
Physically, enhanced ionisation has been attributed to two main causes: a narrower effective-potential barrier for the uphill well due to the presence of a neighbouring, downhill centre, and strongly coupled charge resonant states. This narrowing implies that tunnel ionisation will be enhanced for the uphill well. Typically, a comparison with a companion atom of similar ionisation potential shows an increase of at least one order of magnitude in the ionisation rate. This explanation is strongly based on the quasi-static picture, in which one assumes that the joint influence of the binding potential and the instantaneous driving field form an effective potential barrier. Discussions on the validity of this picture, together with the assumption that the electron tunnels from the uphill well, can be found in \cite{Sheehy1995,Thompson1997,Liu2017}, and its experimental confirmation has been reported in \cite{Wu2012}. Nearly degenerate coupled charge-resonant states that occur for large inter-nuclear separations also facilitate a strong population transfer to the continuum \cite{Zuo1995,Bandrauk1999,Yudin2005}.  On the other hand, for extended systems, multielectron and non-adiabatic effects play an increasingly important role \cite{Erattupuzha2017,Markevitch2003,Markevitch2004}(for a recent review see \cite{Miller2016,Bircher2017}). Loosely speaking, non-adiabaticity implies that the electron probability density as a function of time does not ``follow" the field gradients, and instead may be related to population trapping, resonances, multielectron effects and coupling of different degrees of freedom.  This type of behaviour has also raised  a great deal of debate in the context of the attoclock  \cite{Shafir2012,Landsman2015,Ohmi2015,Liu2019}.

One should note, however, that even simple, one-electron systems with only two wells may exhibit non-adiabatic behaviour. For instance, in \cite{He2008,Takemoto2010}, multiple ionisation bursts that have been identified theoretically do not follow the time profile of the field. It has been shown that the time-dependent density distribution does not follow the time profile of the field. This behaviour was attributed to strongly coupled charge resonant states and the system responding non-adiabatically to time dependent fields \cite{Takemo2011}. In \cite{He2008,Takemo2011}, a phase-space analysis has been performed using Wigner quasiprobability distributions. Thereby, intriguing structures that cycle through momentum space in a quarter of the field cycle have been identified: the momentum gates. As they allow a rapid transfer of population, such gates have been associated to quarter-cycle ionisation bursts, and thus to non-adiabatic following of the time-dependent field \cite{Takemo2011}.

Nonetheless, the physical reasons for this behaviour remain unknown. Furthermore, it is not clear whether momentum gates could occur under a different set of circumstances, such as with static fields, or what role quantum interference and other types of nonclassical behaviour play in this context. A key issue is also how to define nonadiabaticity or nonclassicality of strong-field enhanced ionisation. In order to address these questions, methods that make use of phase space are the ideal tools. A major advantage is that they provide information not only about the electron's initial coordinate, but also about its initial momentum distribution and their subsequent evolution. 

A powerful example are Wigner quasiprobability distributions, which, within the constraints posed by the uncertainty principle, permits the study of position-momentum correlations (for reviews see, e.g.,  \cite{Case2008,Weinbub2018}). Wigner functions have been widely employed in quantum optics and quantum information, but are underused in strong-field and attosecond physics. There are however studies of ionisation \cite{Czirjak2000,He2008,Takemo2011,Zagoya2014}, rescattering \cite{Graefe2012,Baumann2015} and entanglement \cite{Czirjak2013} in this context. 
In strong-field ionisation, seminal studies \cite{Czirjak2000} for a zero-range potential in a static field identified a tail that could be associated with a classical tunnelling trajectory far from the core region, and with tunnel ionisation close to the core.
In subsequent work, we have shown that, for a static field and ionisation from a single centre, the Wigner function follows or is partly contained by classical separatrices \cite{Zagoya2014}. This happens to its bound part, and to the above-stated tail, which closely follows the saddle formed by the interplay of field and the potential, also known as ``the Stark saddle". For longer times, one can also see that this tail moves towards lower momenta, and that the Wigner function exhibits signatures related to the quantum interference of different ionisation events.  Under many circumstances this even means that the electron reaches the continuum with non-vanishing momenta, although it follows an equienergy curve and thus the field gradient. This oscillating behaviour around the separatrix has been identified in other areas of research as the semiclassical limit of the Wigner function's time evolution \cite{Balazs1990}.

In the present work, we have a closer look at the role of quantum interference and non-classical behaviour in enhanced ionisation. Using Wigner quasiprobability distributions, we show that a time-dependent field is not a necessary pre-requisite for momentum gates to form. Instead, their primary cause is quantum interference, which builds a bridge such that there is an abrupt momentum transfer from one ion to the other.  Interference may occur when both wells are occupied, either by upfield-downfield population transfer from a wave packet located in the upfield well, or by an initially delocalised bound state.  Their subsequent evolution is determined by the molecule, and, depending on the molecular and field parameters, may contribute to enhanced ionisation. Typically, near quantum bridges the Wigner quasiprobability distribution exhibits non-classical evolution, which is assessed using the quantum Liouville equation. Such statements are backed by showing that (i) momentum gates may form for static fields, or even in the absence of external electric fields; (ii) they are strongly suppressed if the electronic wave packet is initially localised in the downfield molecular well. In this latter case, the Wigner quasiprobability distribution will tend to follow the classical separatrix, as in the single-atom case.  
We also identify different regimes for which there is population trapping or enhanced ionisation.  This will depend on the driving-field strength and on the inter-nuclear separation. 
 
Our article is organised as follows. After going through the necessary theoretical background in Section \ref{sec:model}, we will give an overview in Section \ref{sec:regions} of the different phase space configurations of our system, as well as their effects on the ionisation rate. Following that, in Section \ref{sec:Wignercat}, we will perform a study of momentum gates for initially delocalised states using both static and field-free systems. In Section \ref{sec:Wignerlocalised} we use initial localised states, both upfield and downfield, to provide more information on the momentum gates and their effect. In section \ref{sec:tevolv}, using both the autocorrelation function and the quantum Liouville equation, we investigate the temporal evolution of our system. In Section \ref{sec:tdep} we study time-dependent fields and their quasi-probability distributions. Finally, in Section \ref{sec:conclusions} we will then close this work with conclusions and discussion. Unless stated otherwise, atomic units are used throughout.
\vspace*{0.5cm}
\section{Model and Method}
\label{sec:model}
In order to facilitate the phase-space analysis and interpretation, in this work we will use a simplified, one-dimensional and one-electron molecular model.  The evolution of the electronic wave packet will be given by the full solution of the time dependent Schr\"odinger equation (TDSE), which, in atomic units, reads
\begin{equation}
i\partial_t \Psi (x,t) = \left(-\frac{1}{2}\frac{d^2}{dx^2}+V(x) + x E(t) \right)\Psi (x,t),
\label{eq:tdse}
\end{equation}
where $\Psi(x,t)$ is the wave function of our system, and $E(t)$ is the external laser field, which is either chosen as a linearly polarised monochromatic wave
\begin{equation}
E(t)=E_0\cos(\omega t),
\label{eq:tdfield}
\end{equation}
of amplitude $E_0$ and frequency $\omega$, or taken to be static, i.e., $E(t)=E$. Here, one may describe the problem in terms of an effective potential 
\begin{equation}
V_{\mathrm{eff}}(x)=V(x)+xE.
\label{eq:Veff}
\end{equation}
The molecular binding potential is given by 
\begin{equation}
V(x)=V_0(x+R/2)+V_0(x-R/2),
\label{eq:potential}   
\end{equation}
where $R$ is the inter-nuclear separation. We choose the potential $V_0$ at each molecular well as the soft-core potential
\begin{equation}
V_0(x)=-\frac{1}{\sqrt{x^2+a}},
\label{eq:softcore}
\end{equation}
where $a=1$ is a softening parameter. This is a widely used model as it is non singular and yet long range \cite{Javanainen1988}.  

The TDSE is then solved numerically using the split-operator \cite{Feit1982} method. Therein, the linear and non-linear propagation steps are split up and treated in the frequency domain and the time domain, respectively.
We will approximate the initial wave function by  Gaussian wave packets 
\begin{equation}
\Psi(x,0)=\langle x|\Psi (0) \rangle = \left( \frac{\alpha}{\pi} \right)^{\frac{1}{4}}\exp \left\{-\frac{\alpha}{2}(x-q_{0})^{2}+ip_{0}(x-q_{0})  \right\}
\label{eq:Psi0}
\end{equation}
of width $\alpha$ centred at vanishing initial momentum $p_0=0$ and initial coordinate $q_0$, or coherent superpositions thereof. 
Specifically, we will consider $q_0=-R/2$ or $q_0=R/2$, in which cases the initial wave functions are given by $\Psi_{\mathrm{down}}(x,0)$ or $\Psi_{\mathrm{up}}(x,0)$, respectively. The widths of $ \Psi(x,0)$ (\(\alpha = 0.5\) a.u.) have been calculated so that the ground-state energy of a single-centre soft-core potential with $a=1.0$ is minimised.

The delocalised wave function is taken to be the symmetric coherent superposition 
\begin{equation}
\Psi_{\mathrm{cat}}(x,0) = \frac{\Psi_{\mathrm{down}} (x,0)+\Psi_{\mathrm{up}}(x,0)}{\sqrt{2\left(1+\mathcal{I}_o\right)}},
\label{eq:statcat}
\end{equation}
where $\mathcal{I}_o$ is the overlap integral
\begin{equation}
\mathcal{I}_o=-\int \Psi^*_{\mathrm{down}} (x,0) \Psi_{\mathrm{up}}(x,0)dx = e^{-R}\left( 1+R+\frac{1}{3} R^{2}\right).
\end{equation}
Within the approximations used, Eq.~(\ref{eq:Psi0}) describes coherent states, while the wavepacket given by Eq.~(\ref{eq:statcat}) is known as a stationary cat state  \cite{schleich2011quantum}.

In order to understand the time dynamics of the wave function, we will be using the time dependent autocorrelation function \(a(t)\) given by
\begin{equation}
a(t) = \int \Psi^* (x,t) \Psi (x,0) dx
\label{eq:autocorrelation}
\end{equation}
as well as the ionisation rate \(\Gamma\) from an initial time  \(t=0\) to a final time \(t=T\), calculated using
\begin{equation}
\Gamma=-\ln\left(\frac{|\mathcal{P}(T)|^{2}}{|\mathcal{P}(0)|^{2}}\right)\frac{1}{T},
\label{eq:ionisation}
\end{equation}
where
\begin{equation}
\mathcal{P}(t)=\int^{+\infty} _{-\infty}\Psi^*(x,t)\Psi(x,t)dx.
\label{eq:normWF}
\end{equation}
Note that, in practice, the limits of the above integral will be finite (typically \(x=-100\) to \(x=100\) a.u.). Furthermore, 
due to irreversible ionisation,  Eq.~(\ref{eq:normWF}) will be less than unity and decrease with time. Thus, it will be a good measure of the probability density that has reached the integration boundaries. This definition of ionisation rate was employed in the seminal paper \cite{Zuo1995}. To minimise reflections or the effect of the absorber, the grid is taken to be twice as large as the `box size' set by the above-stated limits. 

\subsection{Wigner Function}
The Wigner quasi-probability distribution is immensely relevant to this study because, within the constraints dictated by the uncertainty principle, it introduces momentum and position resolution. This allows us to exploits the classical notion of phase space while using a purely quantum mechanical tool. It is given by

\begin{equation}
W(x,p,t)= \frac{1}{\pi} \int_{\infty}^{-\infty}d\xi\Psi^{*}(x+\xi,t)\Psi(x-\xi,t)e^{2ip\xi}, 
\end{equation}

where the position and momentum coordinates are represented by \(x\) and \(p\), respectively.
This function is always real. However, it exhibits both positive and negative values. This, among other features, makes its interpretation as a simple probability distribution difficult \cite{Case2008}. Negative values of the Wigner function are typically associated with non-classicality (see, e.g. \cite{Kenfack2004}).

For the symmetric, delocalised state given by Eq.~(\ref{eq:statcat}), the Wigner function reads
\begin{equation}
W_{\mathrm{cat}}(x,p,0)= \frac{W_{\mathrm{down}}(x,p,0)+ W_{\mathrm{up}}(x,p,0)+W_{int}(x,p,0)}{2\left(1+\mathcal{I}_o\right)} ,
\label{eq:Wigner0cat}
\end{equation}

where
\begin{equation}
W_{j}(x,p,0)=\frac{1}{\pi}\exp\left[\alpha(x \pm R/2)^2-\frac{p^2}{\alpha}\right].
\label{eq:Wignerlr}
\end{equation}

The index $j=\mathrm{down}$ indicates a Wigner function centred at $(q_0,p_0)=(-R/2,0)$, i.e., the downfield well, while $j=\mathrm{up}$ refers to a centre at $(q_0,p_0)=(R/2,0)$, i.e., the upfield well.
The term 

\begin{equation}
W_{int}(x,p,0)= \frac{2}{\pi}\exp\left[-\alpha x^2-\frac{p^2}{\alpha}\right]\cos[pR]
\label{eq:Winterf}
\end{equation}

is peaked at the origin and gives a series of interference fringes parallel to the $x$ axis, whose extrema occur for $p=n\pi/R$. Even and odd values of the integer number $n$ give maxima and minima, respectively. 
If the wave packet is initially localised in the downfield or upfield well, the initial Wigner function will be given by $W_{\mathrm{down}}(x,p,0)$ or $W_{\mathrm{up}}(x,p,0)$, respectively, and the interference term is absent.

A widespread tool to investigate the time evolution of the Wigner function is the quantum Liouville equation \cite{schleich2011quantum}. Explicitly, it reads
\begin{equation}
\left( \frac{\partial}{\partial t} + \frac{p}{M} \frac{\partial}{\partial x}-\frac{dV_{\mathrm{eff}}(x)}{dx} \frac{\partial}{\partial p}\right) W(x,p,t)= Q(x,p,t),
\label{eq:liouville}
\end{equation}
where
\begin{equation}
Q(x,p,t)= \sum_{l=1} ^{\infty} \frac{(-1)^{l}(\hbar/2)^{2l}}{(2l + 1)!} \frac{d^{2l+1}V_{\mathrm{eff}}(x)}{dx^{2l+1}} \frac{\partial^{2l+1}}{\partial p^{2l+1}} W(x,p,t)
\label{eq:quantumcorr}
\end{equation}
are the quantum corrections to the classical Liouville equation. In the classical limit (setting \(\hbar = 0\)),  \(Q(x,p,t) = 0\) and the Wigner function will follow the evolution of a classical particle, i.e., according to the classical Liouville equation. A direct inspection of Eq.~(\ref{eq:quantumcorr}) shows that the quantum corrections vanish for binding potentials up to the second order in $x$. This includes linear potentials such as the interaction Hamiltonian in $V_{\mathrm{eff}}(x)$ and harmonic potentials. 
\section{Phase-space regions}
\label{sec:regions}

Below we identify and analyse the bound and continuum regions in phase space that may occur for different inter-nuclear separations, in the absence and in the presence of external fields. For simplicity, we take the field to be static. This reflects to some extent the instantaneous configurations that will occur for a low-frequency field and will be useful in interpreting the subsequent results.
\begin{figure}[ht]
	\centering
	\includegraphics[width=12cm]{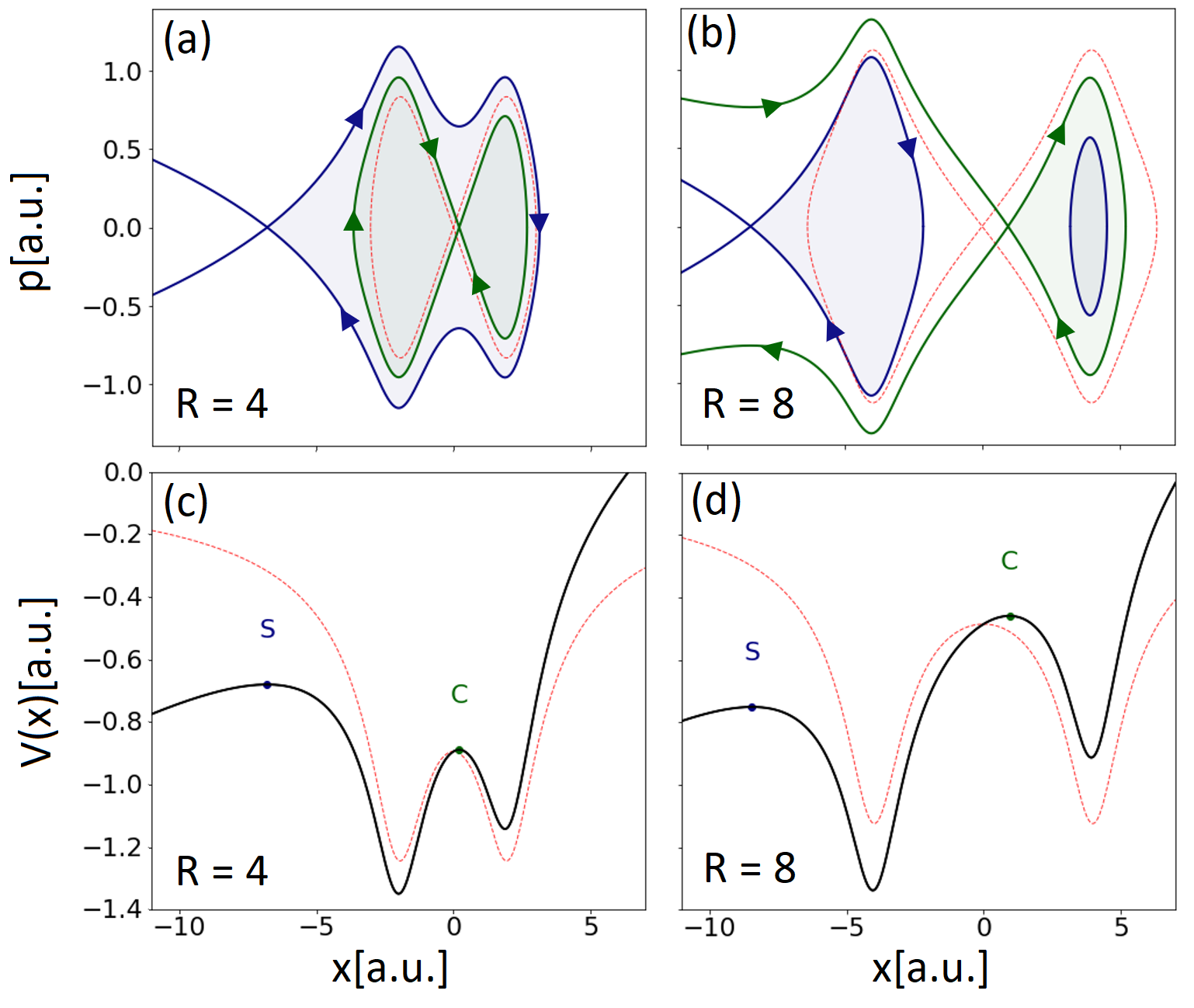}
	\caption{Phase portraits for the one-dimensional homonuclear molecular models described by the softcore potential (\ref{eq:softcore}), using inter-nuclear separations of $R = 4$ a.u. and $R = 8$ a.u. and a static field $E=0.0534.$ a.u. [upper panels], together with the corresponding effective potentials [lower panels]. The Stark and the central saddles are indicated by the labels S and C in the figure, and the field-free separatrices and potentials are given by the dashed red lines. The shaded areas indicate the phase space regions for which the wave packet is bound. The colours of these regions match those of the respective separatrices. }
	\label{fig:bifurcationStudy}
\end{figure}

\begin{figure}[ht]
	\centering
	\includegraphics[width=10cm]{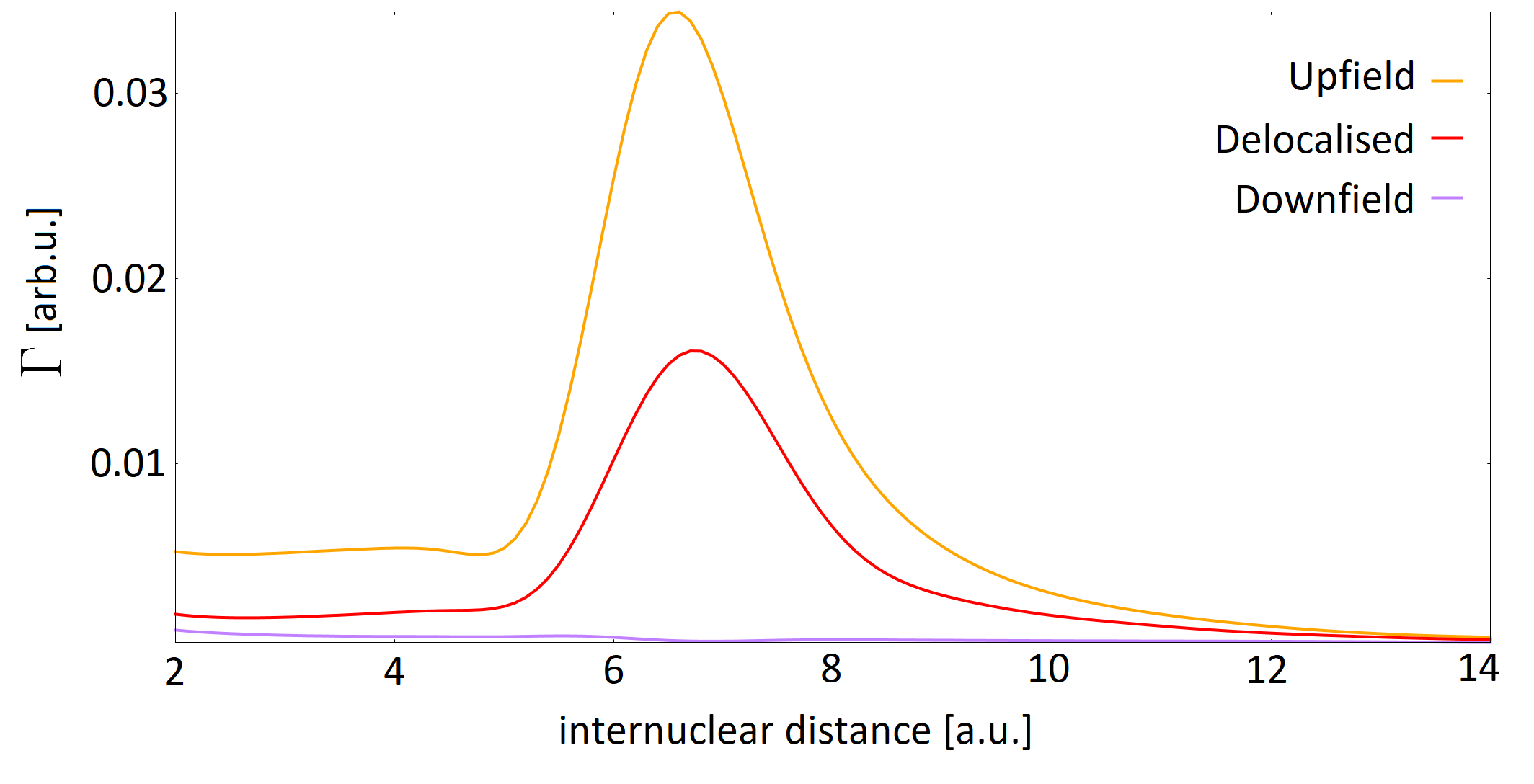}
	\caption{Ionisation rate as a function of the inter-nuclear distance $R$, calculated in a `box' from \(x_{\mathrm{min}}=-100\) a.u. to \(x_{\mathrm{max}}=100\) a.u.,  final time \(T_{end}=150\) a.u. and field strength \(E=0.0534\) (intensity \(I=10^{14} \hspace*{0.1cm}\mathrm{W/cm}^{2})\) using different starting wave packets: delocalised (red), localised upfield (orange) and localised downfield (purple). The vertical line indicates the inter-nuclear separation for which the phase-space configuration changes.}
	\label{rateOverDistance}
\end{figure}

These regions are illustrated in Fig.~\ref{fig:bifurcationStudy} for several parameter ranges, for the potential described by Eq.~(\ref{eq:softcore}). 
In the field-free case (see dashed lines in Fig.~\ref{fig:bifurcationStudy}), the separatrices are symmetric with regard to a central saddle. They form a closed curve around the two molecular centres, consisting of homoclinic trajectories. 
A non-vanishing field breaks this symmetry and causes the appearance of two separatrices, which are asymmetric with regard to the a reflection about $x=0$. It also gives rise to an additional saddle on the left, the Stark saddle. Hence, in the cases most relevant to us, there are two saddles and two centres, whose interplay will influence the ionisation dynamics. 
This leads to two different phase space regimes, shown in Figs.~\ref{fig:bifurcationStudy}(a) and (b). The energy difference between the Stark saddle and the central saddle is an important parameter for determining whether enhanced ionisation may or may not occur. 
If the downfield potential maximum is higher in energy than the upfield maximum [left panels in Fig.~\ref{fig:bifurcationStudy}], then the separatrix associated with the Stark saddle (in blue) encapsulates entirely the homoclinic separatrix related to the central saddle (in green). Thus, even if an electronic wave packet tunnels through the central barrier, it would be trapped in the downhill centre. This means that it would still need to tunnel through a wider Stark saddle in order to reach the continuum.
In contrast, if the energy of the central saddle is higher than that of the Stark saddle [right panels in Fig.~\ref{fig:bifurcationStudy}], the electron would only need to tunnel through the central saddle to reach the continuum.
The optimal scenario occurs if the energy of the central saddle is high enough to allow direct ionisation into the continuum, but still leads to an effective potential barrier narrower than that of a single atom.
In the presence of the field, an initial wave packet described by $\Psi_{\mathrm{up}}(x,0)$ or $\Psi_{\mathrm{down}}(x,0)$ is centred in the upfield or the downfield well, respectively, while the delocalised wave packet $\Psi_{\mathrm{cat}}(x,0)$ occupies both.

The ionisation rate as a function of the inter-nuclear distance, plotted in Fig.~\ref{rateOverDistance}, shows the effect of the different phase-space configurations. If the initial wave packet is  delocalised, or centred around the upfield potential minimum, a very strong peak is present. The ionisation rate starts to increase dramatically when the inter-nuclear distance reaches the value at which the phase-space configuration changes. For this critical value of R, which, for the chosen external field, is \(R_c=5.2\) a.u. (see thin vertical black line), the outer separatrix ``opens" and no longer traps the wave packet. The ionisation rate then reaches a peak at \(R=6.8\) a.u., and drops for larger inter-nuclear separations. This happens because the two centres become further apart, thus hindering tunnel ionisation via the central saddle. If a downfield initial wave packet $\Psi_{\mathrm{down}}(x,0)$ is taken, the ionisation rate is suppressed by two orders of magnitude.

This, along with the fact that the ionisation rate with an upfield wave packet is about double that of a delocalised wave packet, suggests that around the opening of the separatrix  ionisation comes mainly from the upfield population. 
A comparison with Fig.~\ref{fig:bifurcationStudy} shows that this corresponds exactly to the hypothesised point of maximum enhancement, where the energy of the saddle is high enough for  the tunnelling electron not to be trapped by the downfield centre (e.g. after \(R_c=5.2\) a.u.) but low enough for the effective potential barrier to be narrower than that of a single atom (e.g. before \(R=7\) a.u.).

\section{Momentum gates for initial delocalised states}
\label{sec:Wignercat}
\begin{figure}[H]
	\centering
	\includegraphics[width=13.0cm]{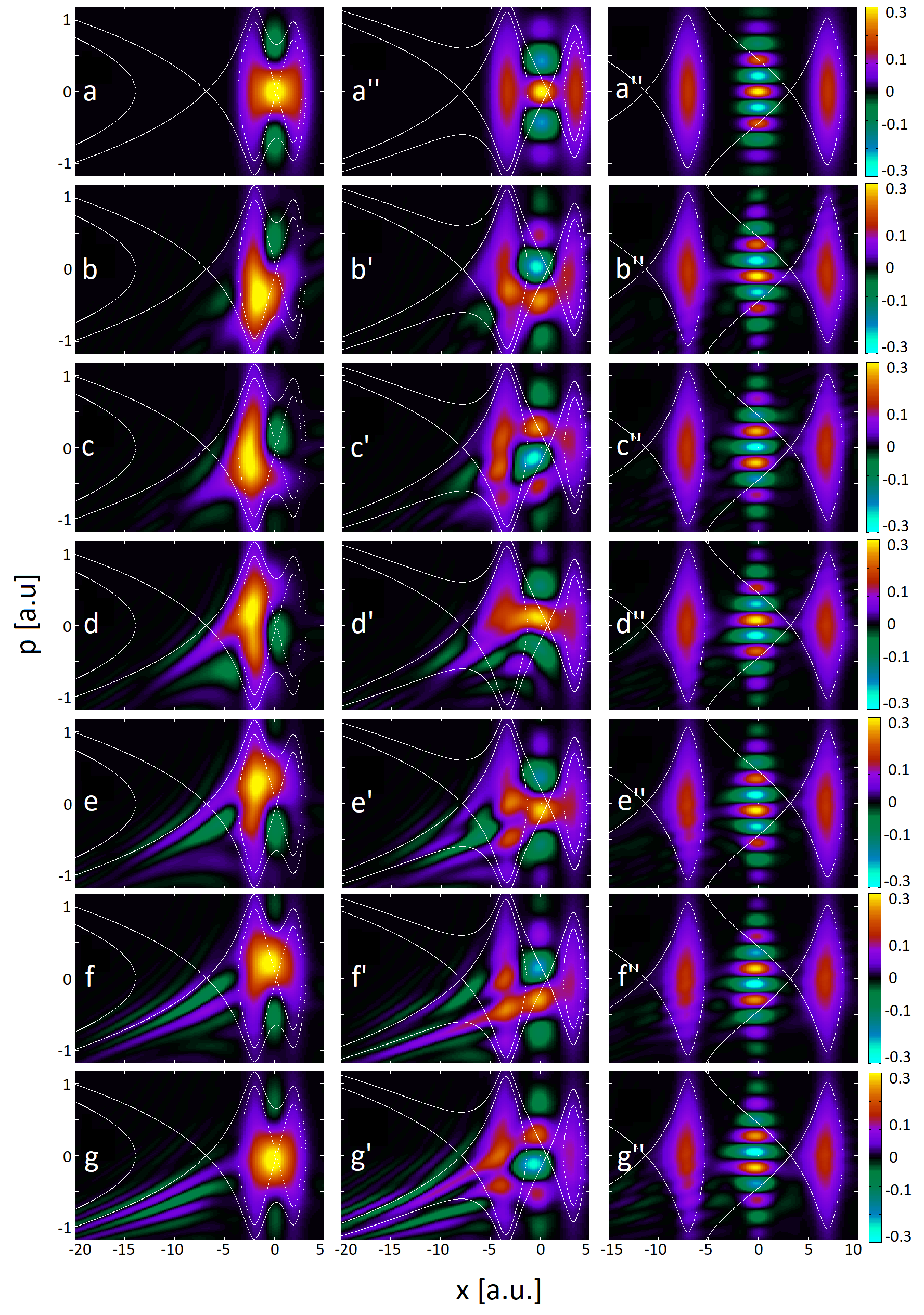}
	\caption{Wigner quasi probability distribution at different instants of time, calculated for a model $H_2^+$ molecule in a static laser field of strength \(E=0.0534\) a.u. (intensity \(I=10^{14}  \mathrm{W/cm}^{2})\) using an initially delocalised (cat) state given by Eq.~(\ref{eq:statcat}), with $\alpha=0.5$. In the left, middle and right columns, the inter-nuclear separation is taken as \(R=4\) a.u., \(R=6.8\) a.u. and \(R=14\) a.u., respectively. The temporal snapshots are given from top to bottom. Panels (a), (a$'$) and (a$'' $) [first row] have been calculated for $t = 0$ a.u., panels (b), (b$'$) and (b$'' $) [second row]  for $t = 8$ a.u., panels (c), (c$'$) and (c$'' $)  [third row] for $t = 12$ a.u., panels (d), (d$'$) and (d$'' $)  [fourth row] for $t = 16$ a.u., panels (e), (e$'$) and (e$'' $)   [fifth row] for $t = 20$ a.u., panels (f), (f$'$) and (f$'' $) [sixth row] for $t = 24$ a.u.,  panels (g), (g$'$) and (g$'' $)  [seventh row] for $t = 30$ a.u. The thin white lines in the figure give the equienergy curves (including the separatrices). }
	\label{fig:wignerFull}
\end{figure}

Using the Wigner quasiprobability distributions, we provide a more detailed explanation for this enhancement and its causes. In Fig.~\ref{fig:wignerFull}, we plot such distributions for different inter-nuclear separations computed using initial delocalised states $\Psi_{\mathrm{cat}}(x,0)$.  
The inter-nuclear distances used in the left, centre and right panels, respectively, have been chosen such that (i) the two separatrices are nested and closed (\(R=4.0\) a.u.), (ii) the outer separatrix has just opened (\(R=6.8\) a.u.), and (iii) the separatrix including the central saddle is completely open and the two centres are well separated (\(R=14.0\) a.u.).

The initial Wigner functions $W(x,p,0)$ are given by Eq.~(\ref{eq:Wigner0cat}), and behave as predicted, with Gaussian shaped quasiprobability densities centred at the origin \((x,p)=(0,0)\) and at each potential well $(x,p)=(\pm R/2,0)$. There are also interference fringes near the central saddle, with extrema at $(x,p)=(0,n\pi)$, which become finer for larger values of $R$. This pattern is less distinguishable for small inter-nuclear distances [Fig.~\ref{fig:wignerFull}(a)] due to the strong overlap of the Gaussians that form $W(x,p,0)$, but becomes clearer as this overlap decreases [see, e.g., Fig.~\ref{fig:wignerFull}(a$'$)]. For $R=14$ a.u., the central fringes and the Gaussians located in the uphill and downhill potentials are very well defined, with little overlap.

With time propagation, the Wigner functions become asymmetric, flowing from the upfield to the downfield well in the molecule. Semiclassically, the expected behaviour is that the Wigner quasiprobability density follows the classical separatrices and form a tail that can be associated with over-the-barrier or tunnel ionisation, as well as with an oscillatory behaviour around the separatrix. This behaviour was first predicted in \cite{Balazs1990}, and reported in \cite{Czirjak2000,Zagoya2014} in the context of strong-field ionisation. It can be seen clearly on the left of the downfield well, for $t\geq 20$ (third to last row in Fig.~\ref{fig:wignerFull}).

However, there are very peculiar features that do not follow the separatrices, form at much earlier times and occur in the region around the Stark saddle [see, for instance, Fig.~\ref{fig:wignerFull}(b), (b$'$) and (b$''$)]. They consist of a strong quasiprobability flow from one centre to the other.
This transfer occurs  mostly along lines of approximately vanishing phase-space slope, i.e., of nearly constant momentum, and are the momentum gates reported in Refs.~\cite{Takemoto2010,Takemo2011}.
They are visible for the whole range of inter-nuclear separations in Fig.~\ref{fig:wignerFull}, although they manifest themselves in different ways. The interference fringes around the central saddle act as a quantum bridge and facilitate this transfer for positive quasiprobability densities.
The flow is significant if the overlap of the left and right peaks with the central interference structure is large, as shown in the left and central columns of Fig.~\ref{fig:wignerFull}.  Therein, the Wigner function exhibits a clockwise movement, whose period depends on the inter-nuclear separation. Figs.~\ref{fig:wignerFull}(b) and (b$'$) show the start of this motion, with a strong right-left flow for momenta below that of the central saddle. This momentum gate then moves upward in phase space until a subsequent bridge is established, and the bulk of the Wigner function is transferred back.

Furthermore, the presence or absence of enhanced ionisation is directly linked to the interplay of the semiclassical tail and the population transferred via the quantum bridge. If the separatrices are nested (left panels of Fig.~\ref{fig:wignerFull}), the bulk of the quasiprobability distribution remains trapped by the inner separatrix and tunnels back to the upfield centre. This trapping can be clearly seen on the left-hand side of  Fig.~\ref{fig:wignerFull}(b), in which the inner separatrix hinders the Wigner function to reach the Stark saddle. Significant tunnelling via this saddle may only occur after population has built up in the downfield centre, at later times [see Fig.~\ref{fig:wignerFull}(d) and (e)]. There is also some ``spilling" of the Wigner function for larger absolute values of $p$,  when the two separatrices become close in phase space. This spilling can be seen at the bottom  of  Figs.~\ref{fig:wignerFull}(b) and (e), but it is not a highly probable pathway. 

If, on the other hand, the two separatrices are no longer nested, population trapping will no longer occur. Thus, the tail near the Stark saddle will build up already for $t=8$ a.u.[Fig.~\ref{fig:wignerFull}(b$'$)]. This will add up to the contributions from the tail that forms for higher absolute values of momenta, when the separatrices' energies are close [Fig.~\ref{fig:wignerFull}(c$'$)]. Figs.~\ref{fig:wignerFull}(d$'$) to (g$'$) show that, for later times, direct transfer via the quantum bridges will feed into both tails, which will cause enhanced ionisation. Particularly striking is Fig.~\ref{fig:wignerFull}(f$'$), which  shows a direct quasiprobability leak from the uphill centre to the continuum via the quantum bridge, at a higher energy than that determined by the Stark saddle.

For larger values of $R$, there is far less quasiprobability transfer, but the bridges can be clearly seen due to the three phase-space regions of the Wigner functions being well separated (see right columns in Fig.~\ref{fig:wignerFull}). For instance, for \(R=14\) a.u., at $t=8$ a.u. [Figs.~\ref{fig:wignerFull}(b$''$)], a horizontal bridge forms near zero momentum, and left-right population transfer occurs. Subsequently, the central fringes move downwards. If a negative (positive) quasiprobability density is located near the central saddle, the bridges are weakened (strengthened) [see e.g., Figs.~\ref{fig:wignerFull}(c$''$) and (f$''$) in contrast to Figs.~\ref{fig:wignerFull}(b$''$), (d$''$), and (g$''$)]. If the central bridge is weakened, other bridges may occur for higher, albeit constant momenta.

\begin{figure}[ht]
	\centering
	\includegraphics[width=9cm]{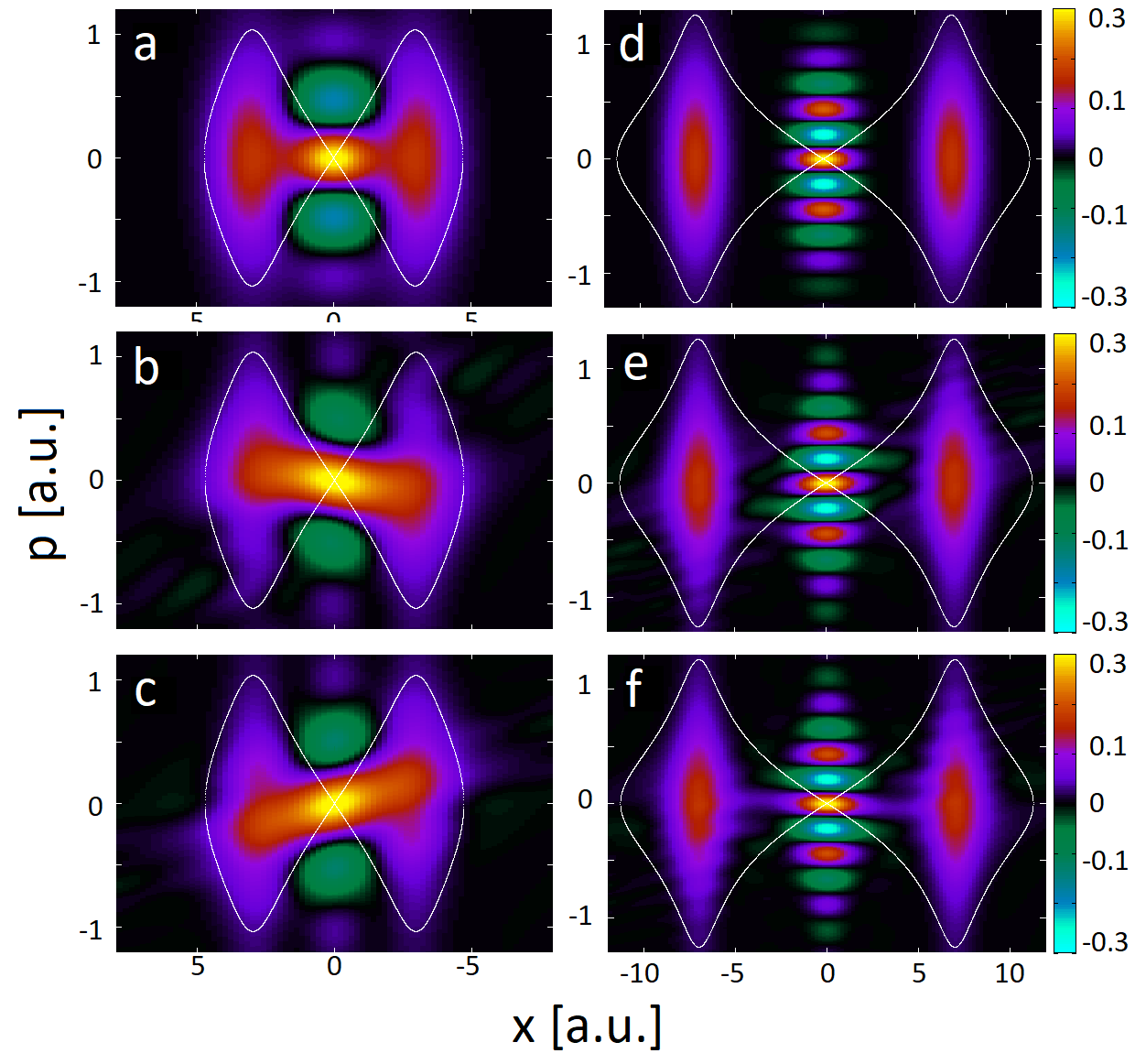}
	\caption{ Wigner function computed using the same initial state as in Fig.~\ref{fig:wignerFull}, but considering  a field-free $H_2^+$ model molecule.  The left and the right columns have been calculated for inter-nuclear separations of \(R=6\) a.u. and \(R=14\) a.u., respectively.  The labels (a) and (d) refer to \(t=0\),  (b) to \(t=5\), (c) to \(t=12\), (e) to \(t=15\), and (f) to \(t=20\).}
	\label{fig:fieldFree}
\end{figure}

The above-mentioned cyclic evolution of the momentum gates is even present for the Wigner function of a field-free $H_2^+$ molecule, shown in Figure~\ref{fig:fieldFree}.  
For smaller inter-nuclear distances, e.g. \(R=6\) a.u., the quasiprobability density ``wobbles" from a positive to a negative gradient. There is a flow  from one centre to another, facilitated by the quasiprobability maximum at \(p=0\). The frequency of this change in the gradient increases with the  inter-nuclear distance. 
For larger inter-nuclear distances, where the overlap region is separated from the two centres, the flow from one centre to another is a lot weaker and is characterised by links between different interference fringes.  First, a simultaneous flow occurs in the positive (negative) momentum region towards the upfield (downfield) centre, see Fig.~\ref{fig:fieldFree}(d). Following that, a bridge between the potential well populations and the saddle population emerges near $p=0$, see Fig.~\ref{fig:fieldFree}(e).  

\section{Momentum gates for initially localised Wigner functions}
\label{sec:Wignerlocalised}
\begin{figure}[ht]
	\centering
	\includegraphics[width=13cm]{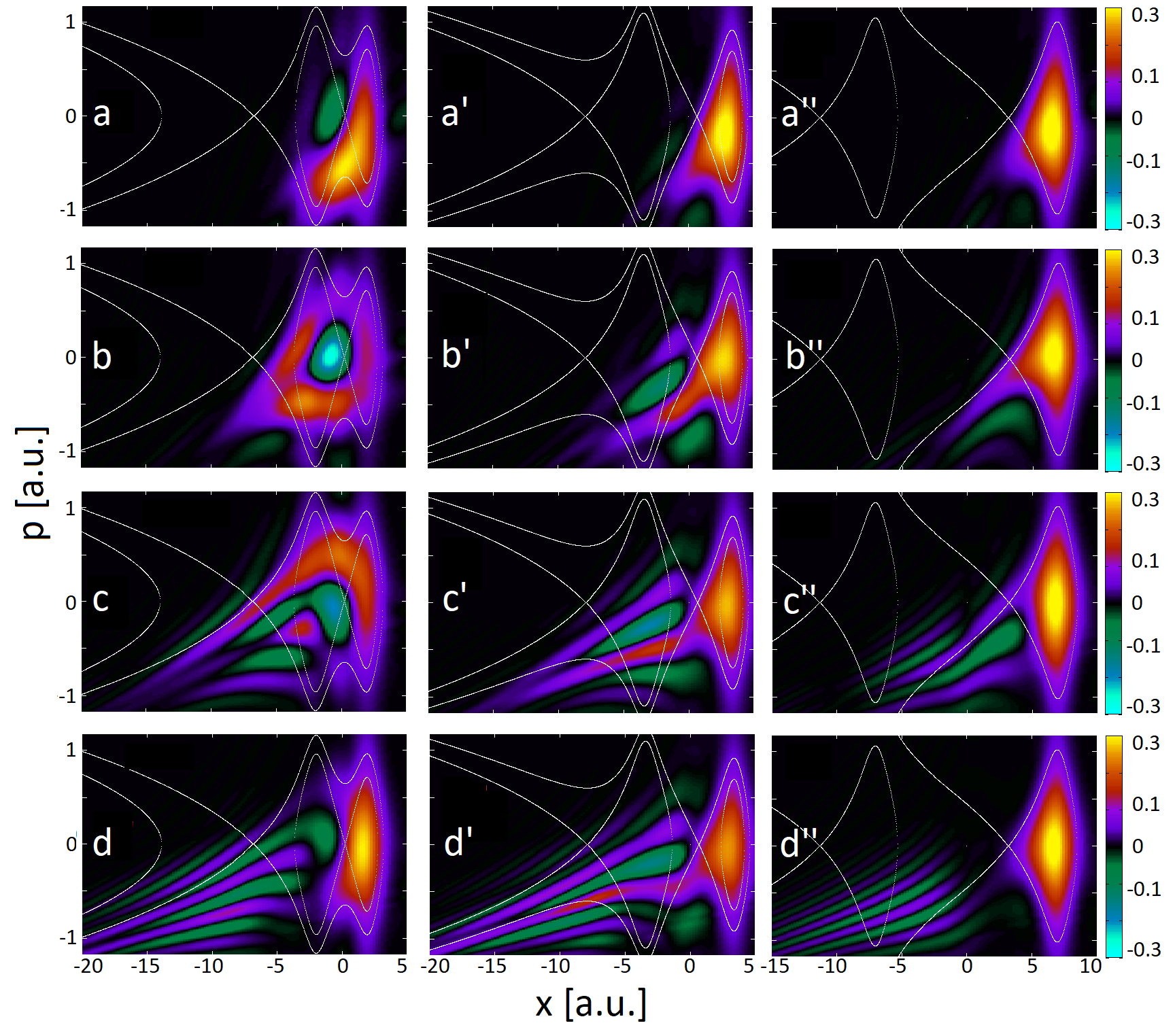}
	\caption{Wigner quasi probability distribution computed for a H$_2^+$ molecule in a static laser field of strength \(E=0.0534\) using a Gaussian initial wave packet $\Psi_{\mathrm{up}}(x,0)$ of width $\alpha=0.5$ centred around the upfield potential well. The left, centre and right columns correspond to the inter-nuclear separations \(R = 4\) a.u., \(R = 6.8\) a.u. and \(R = 14\) a.u., respectively.  The first, second, third and fourth row have been calculated for $t=6.0$ a.u. [panels (a), (a$'$) and (a$''$)], $t= 12.0$ a.u. [panels (b), (b$'$) and (b$''$)], $t=20.0$ a.u. [panels (c), (c$'$) and (c$''$)] and $t=30.0$ a.u. [panels (d), (d$'$) and (d$''$)].}
	\label{fig:wignerUpfield}
\end{figure}

To expand on the roles of the quantum bridges and their  cyclical motion, in the two subsequent figures we employ a similar system, but with localised initial wave packets. In this case, the central interference fringes in the initial Wigner function given by Eq.~(\ref{eq:Winterf}) are absent.
For a wave packet $\Psi_{\mathrm{up}}(x,0)$ placed at the upfield potential well, the dynamics and nature of the bridges are different from those observed in the delocalised case. If the potential wells are not close enough, the quantum bridge does not form and there is no enhanced ionisation. We see this for $R = 14$ a.u. (right panels of Fig.~\ref{fig:wignerUpfield}), where the tail marking the escape path follows the separatrix associated with the uphill centre [Fig.~\ref{fig:wignerUpfield}(b$''$)]. Subsequently, it deviates from this curve when the escaping electron is slowed down by the downfield centre [Fig.~\ref{fig:wignerUpfield}(d$''$)], but no shortcut to the continuum is provided. This is radically different from the \(R = 4\) a.u. case, displayed in the left column of Fig.~\ref{fig:wignerUpfield}, where both the momentum gates and the clockwise motion of the Wigner function are present. Because the separatrices are nested, the population flowing to the downfield centre via the momentum gate is trapped. It then travels back via a positive momentum gate to the upfield centre. Finally, for $R=6.8$ a.u. (middle column of Fig.~\ref{fig:wignerUpfield}), we present the optimal configuration. Indeed the potential wells are close enough to allow the creation of the quantum bridge. However, because the separatrices are open, the population escapes directly through the semiclassical path, following the separatrices, and does not flow back to the upfield centre. The quantum bridges provide a ``shortcut" to several  pathways for the quasiprobabilities to reach the continuum. A clear example is provided in Fig.~\ref{fig:wignerUpfield}(d$'$), which shows a tail starting along the central saddle and being guided by the quantum bridge towards the Stark saddle. Escape happens via several equienergy curves, not only the inner separatrix. 

\begin{figure}[ht]
	\centering
	\includegraphics[width=13cm]{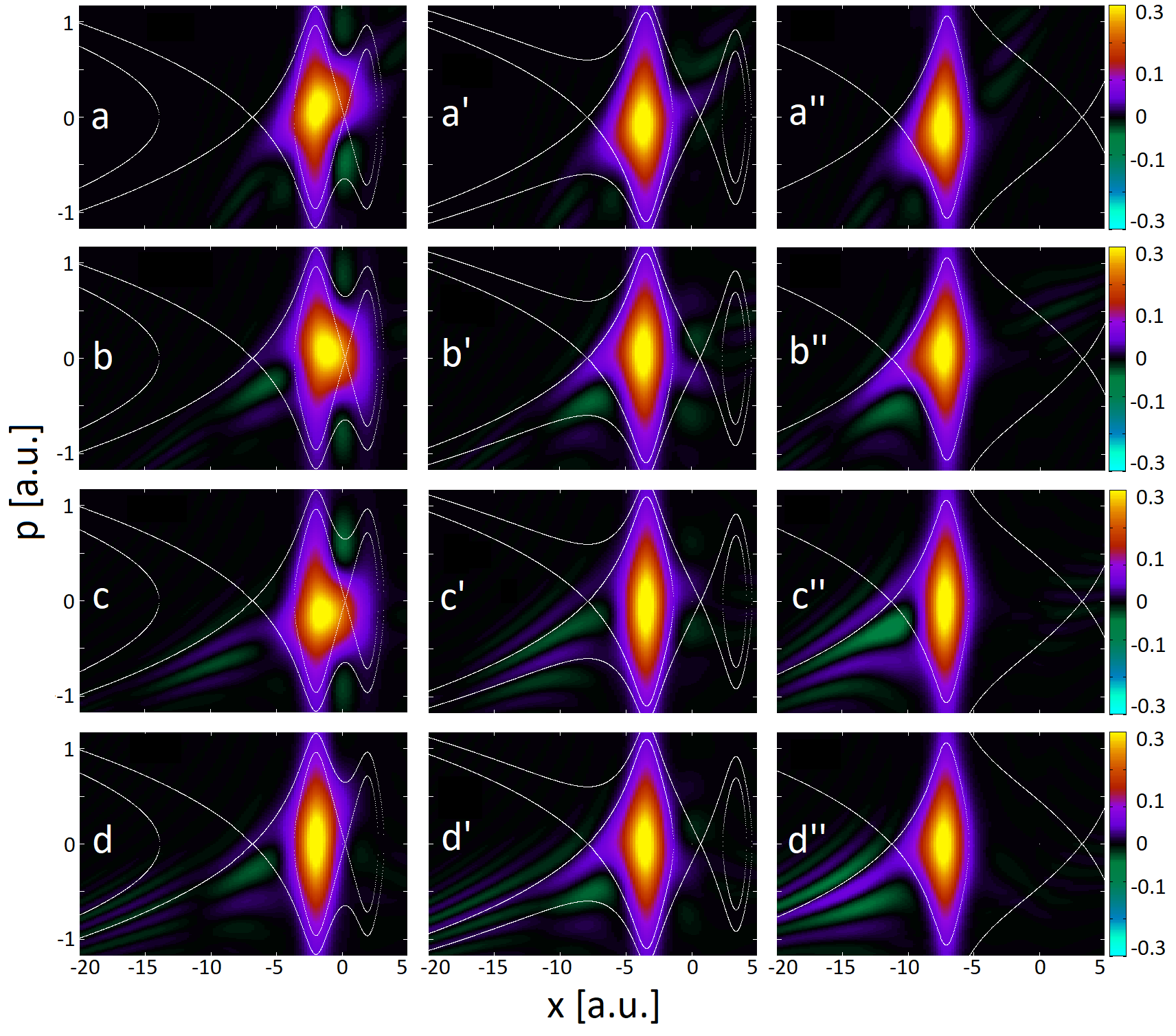}
	\caption{Wigner quasi probability distribution computed for a H$_2^+$ molecule in a static laser field of strength \(E=0.0534\) using a Gaussian initial wave packet $\Psi_{\mathrm{down}}(x,0)$ of width $\alpha=0.5$ centred around the downfield potential well. The left, centre and right columns correspond to the inter-nuclear separations \(R = 4\) a.u., \(R = 6.8\) a.u. and \(R = 14\) a.u., respectively.  The first, second, third and fourth row have been calculated for $t=6.0$ a.u. [panels (a), (a$'$) and (a$''$)], $t= 12.0$ a.u. [panels (b), (b$'$) and (b$''$)], $t=20.0$ a.u. [panels (c), (c$'$) and (c$''$)] and $t=30.0$ a.u. [panels (d), (d$'$) and (d$''$)].}
	\label{fig:wignerDownfield}
\end{figure}

In this context, it is noteworthy that the dynamics and the nature of the bridges are different from those observed in the delocalised case. Whilst we do see multiple tails enhanced by tunnelling from the upfield well, the clockwise motion observed in Fig.~\ref{fig:wignerFull} is much less clear. This quasiprobability transfer from one centre to the other, displayed in the last row of the figure, is only obvious when the separatrices are nested, i.e., for $R<R_c$ [see left panels in Fig~\ref{fig:wignerUpfield}]. In this case, population trapping will hinder enhanced ionisation. However, for $R>R_c$, there will be no such trapping. Furthermore, the clockwise motion of the Wigner function will be strongly suppressed, with no feedback loop to the upfield centre. This makes the upfield localised configuration more efficient for enhanced ionisation than using a delocalised initial state. 

In Fig.~\ref{fig:wignerDownfield}, we present the Wigner probability distributions for an initial downfield wavepacket. In this case, the previously observed quantum bridges are absent throughout and the escape pathway mainly follows that of a single atom \cite{Zagoya2014}, i.e., along the separatrix determined by the Stark saddle. 
Only if the two separatrices are nested and energetically close, i.e., for $R<R_c$, there is some upfield quasiprobability flow, as shown in the first column of the figure. This is however not sufficient to form a bridge between both centres. Nested separatrices mean that ionisation will be strongly suppressed, which can be inferred by the very faint tails of the Wigner function in the continuum region. 

\section{Temporal evolution}
\label{sec:tevolv}
\begin{figure}[ht]
	\centering
	\includegraphics[width=9cm]{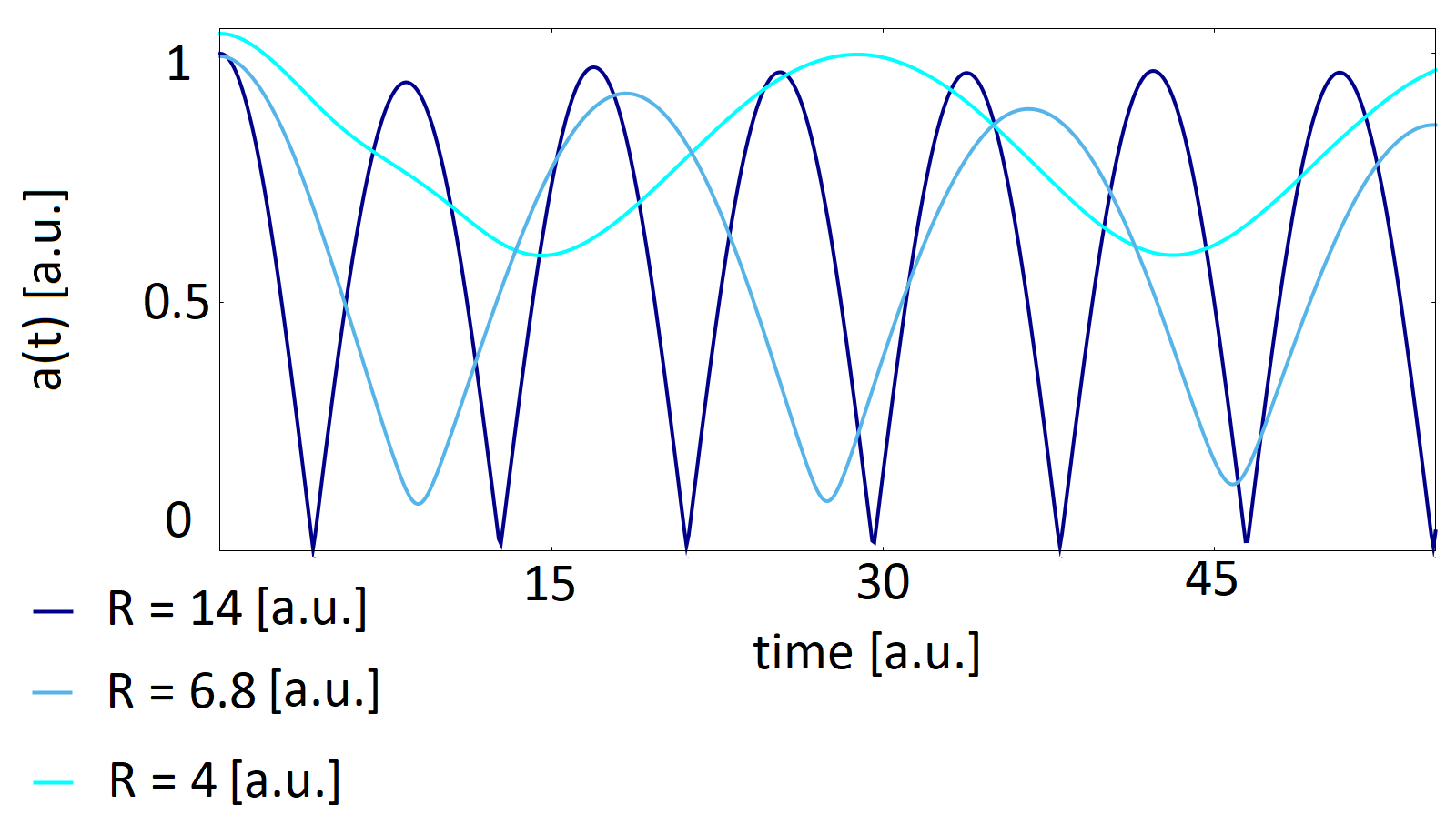}
	\caption{Absolute value of the auto-correlation function of a $H_{2}^{+}$ molecule in a static laser field of intensity \(E = 0.0534\)a.u. using a delocalised starting wave packet with inter-nuclear distances of \(R = 4\) a.u. (cyan), \(R = 6.8\) a.u. (dark blue) and \(R = 14\) a.u. (light blue).}
	\label{fig:autocorrelation}
\end{figure}

We can now focus on the cyclical evolution of the momentum gates, present even in a field-free system (e.g., Fig~\ref{fig:fieldFree}).  This evolution can be quantified directly using autocorrelation functions. In Fig.~\ref{fig:autocorrelation}, we plot their absolute values computed for an initial delocalised state $\Psi_{\mathrm{cat}}(x,0)$ and the same parameters used in  Fig.~\ref{fig:wignerFull}.  As the quasiprobablity distribution shifts towards the downfield centre, the autocorrelation function decreases before increasing again as the population returns to the upfield centre. 
It then reaches its starting position, completing a period of \(T = 29\) a.u. for \(R = 4\) a.u., \(T = 18.3\) a.u. for \(R = 6.8\) a.u. and \(T = 8.4\) a.u. for \(R = 14\) a.u. This is confirmed by comparing the  Wigner functions for \(R = 4\) a.u. at \(t = 0\) and \(t=30\) [Figs. ~\ref{fig:wignerFull}(a) and (g), respectively]. The quasiprobability densities in the bound phase-space region are nearly identical. Significant differences between both Wigner functions occur only in the continuum region, for which there are tails along equienergy curves in the latter time. 

\begin{figure}[ht]
	\centering
	\includegraphics[width=9cm]{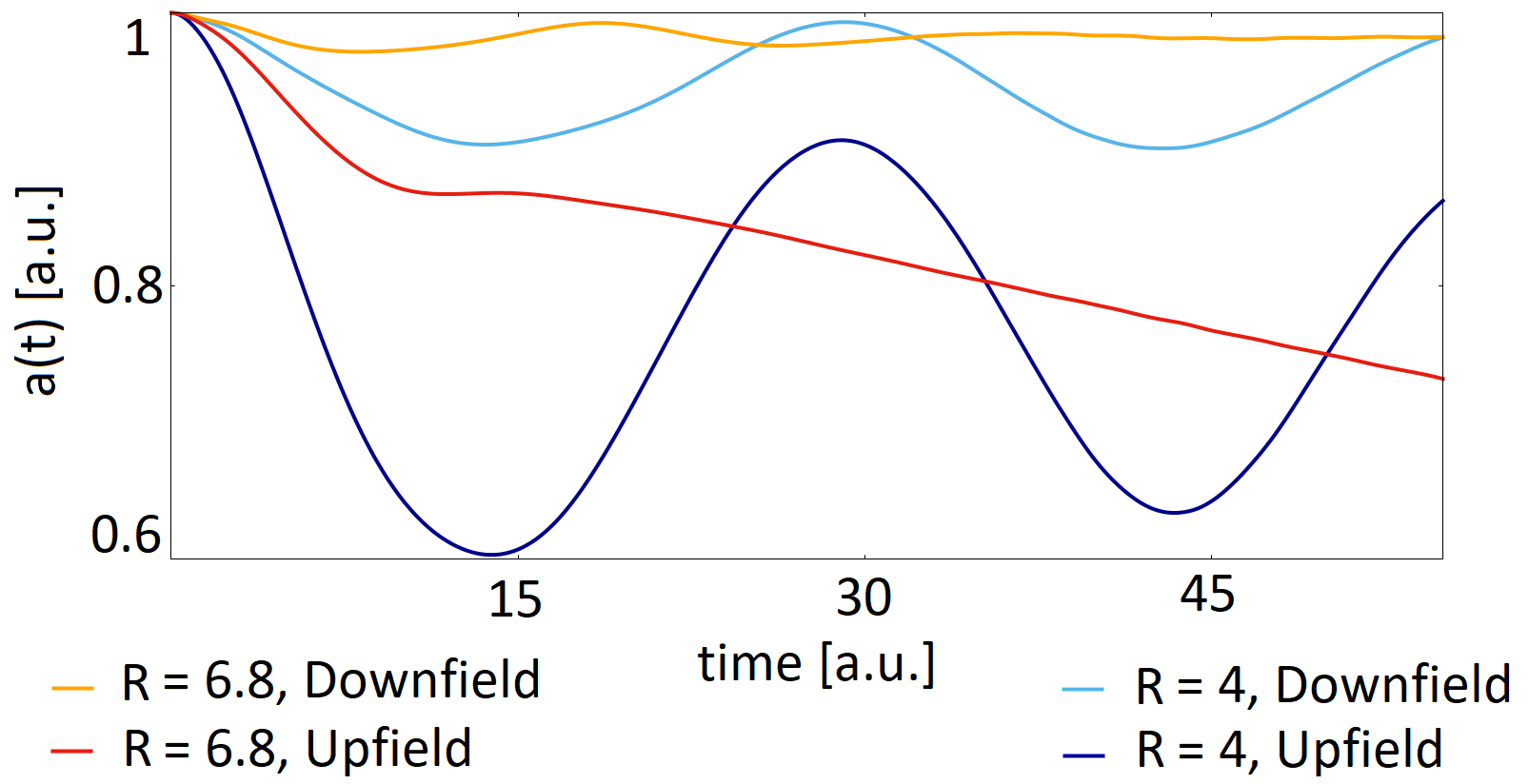}	
	\caption{Absolute value of the auto-correlation function of a \(H_2^+\) molecule in a static laser field of intensity \(E = 0.0534\) a.u. using a localised starting wavepacket with inter-nuclear distances of \(R = 4\) a.u. in blue (upfield) and cyan(downfield), \(R = 6.8\) a.u. in red (upfield) and orange (downfield).}
	\label{fig:autocorrelationUpDown}
\end{figure}

In contrast, Fig.~\ref{fig:autocorrelationUpDown} shows that, for initially localised wavepackets, this clockwise motion is not always present and the autocorrelation function may not be periodic. In this case, periodic behaviour only occurs for $R<R_c$, i.e., when the separatrices are nested and the outer separatrix causes the population to be trapped. For instance, for \(R=4\) a.u., despite starting in a localised state, the autocorrelation function oscillates with a similar frequency both in the upfield and downfield case.  The amplitude of that oscillation however is a lot greater if an initial upfield wavepacket $\Psi_{\mathrm{up}}(x,0)$ is taken, and resembles that obtained using a delocalised starting wavepacket. %Because of the nested separatrices,
This cyclic behaviour however changes if the separatrices are no longer nested, i.e., $R>R_c$.  For the optimal inter-nuclear distance \(R = 6.8\) a.u., if the initial wavepacket is located downfield, the oscillation is very faint, practically absent. For an initial upfield wavepacket $\Psi_{\mathrm{up}}(x,0)$, the autocorrelation function decays practically monotonically. Thus, the electron escapes out of the upfield potential while its return is blocked, maximising ionisation enhancement

In order to understand the non-trivial, sometimes periodic behaviour discussed above, to first approximation one may resort to classical arguments. For that purpose, it is helpful to consider the periodic bound orbits of a classical electron moving in the effective potential (\ref{eq:Veff}) that arises in the presence of a static field. Such orbits have well-defined energies, and their dynamics are governed by Hamilton's equations. This explains the clockwise evolution of the momentum gates, which agree with the directional arrows in Fig~\ref{fig:bifurcationStudy}.

Periodic motion implies closed orbits, whose classical period $T_{cl}$ can be computed with the integral
\begin{equation}
	T_{cl} = \oint_{sep} \frac{1}{p(x)} dx,
	\label{eq:classEstimates}
\end{equation}
over a closed equienergy curve in phase space. As we are interested in the highest possible energy a bound electron may have, we consider
the closed path to be along the separatrix. These estimates can be used as long as the inter-nuclear distance is small enough so that the phase space configuration contains homoclinic trajectories which circle both centres [see Fig.~\ref{fig:bifurcationStudy}(a)]. This condition holds for inter-nuclear distances $R<R_c$, since in that case  the separatrices are nested. 

We obtain a maximum and minimum value for $T_{cl}$ following the path along the outer and inner separatrices, respectively.
\begin{table}[ht]
	\centering
	\begin{tabular}{|c||c|c||c|c||c|c|}
		\hline \hline 
		\multicolumn{1}{|p{2.5cm}||}{\textbf{Initial state}} & \multicolumn{2}{|c||}{$R = 4$}&
		\multicolumn{2}{|c||}{$R = 4.5$}& \multicolumn{2}{|c|}{$R = 5$}\\
		\cline{2-7}
		& $T_{cl}$ & $T_{q}$ & $T_{cl}$ & $T_{q}$ & $T_{cl}$& $T_{q}$\\
		\hline
		$\Psi_{up}$   &  31.9 - 38.7  & 29.0 & 39.5 - 51.4 & 29.8 & 50.1 - 51.9 & NA\\
		$\Psi_{cat}$& 31.9 - 38.7 & 28.8 &  39.5 - 51.4  & 28.3 & 50.1 - 51.9  & 26.1\\
		$\Psi_{down}$& 38.7 - 31.9 & 29.0 & 39.5 - 51.4  & 29.7 &  50.1 - 51.9 & NA\\
		\hline
	\end{tabular}
	\caption{Comparison of the period $T$ obtained from the classical estimates (\(T_{cl}\)) given by  Eq.~(\ref{eq:classEstimates}) and from the absolute value of the auto-correlation function (\(T_{q}\)) of a \(H_2^+\) molecule in a static laser field of strength \(E=0.0534\) a.u. (intensity \(I=10^{14}  \mathrm{W/cm}^{2})\) computed for with different inter-nuclear distances of \(R = 4\) a.u., \(R = 4.5\) a.u. and \(R = 5\) a.u. using different initial states (delocalised (cat), localised upfield and downfield) with width $\alpha=0.5$ a.u.}
	\label{fig:totalTable}
\end{table}
Using the period \(T_{q}\) obtained from the autocorrelation function of the same system, we can compare the classical evolution estimates to the empirical results. 
These results are summarised in Table \ref{fig:totalTable}, with the classical estimates being far above the empirical values in all cases. Furthermore, the period \(T_{cl}\) increases with the inter-nuclear distance, while the opposite trend is observed for \(T_{q}\) (see also the discussion of Fig.~\ref{fig:autocorrelation}, for a wider range of internuclear separations). Throughout, the values of \(T_{q}\) obtained for initial delocalised states are slightly lower. 
 At \(R = 5\)a.u., while the separatrices are still nested, the phase space configuration is very close to the bifurcation discussed in Sec.~\ref{sec:regions} that will lead to an open outer separatrix. 
 If the initial wave function is in a localised initial state, the oscillation no longer takes place and the system resembles that of the open configuration, seen in the middle column of Figs.~\ref{fig:wignerUpfield} and ~\ref{fig:wignerDownfield}, and in Fig.~\ref{fig:autocorrelationUpDown}. Classically, an oscillation is still expected as the orbits remain closed. All this indicates that the evolution of the Wigner function is non-classical.

\begin{figure}[ht]
	\centering
	\includegraphics[width=13cm]{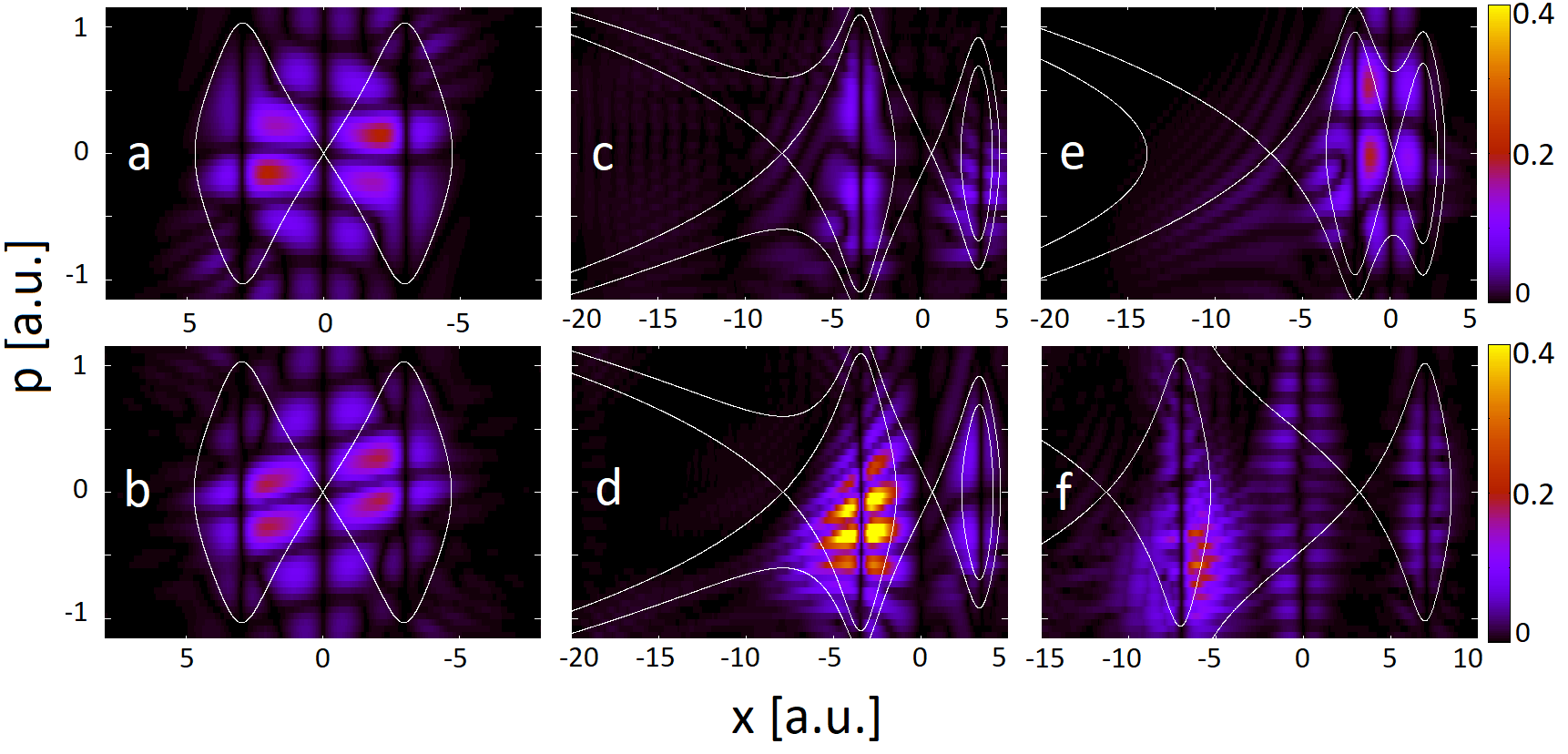}
	\caption{Phase space map of the quantum corrections \(Q(x,p,t)\) [eq.(~\ref{eq:liouville})] of a \(H_2^+\) molecule. (a) and (b) are a field-free delocalised wave packet of inter-nuclear distance \(R = 6\) a.u. at times \(t = 5\) and \(t = 13\) respectively. (c) and (d) both use an \(H_2^+\) molecule in a static laser field of strength \(E=0.0534\) a.u. (intensity \(I=10^{14}  \mathrm{W/cm}^{2})\), inter-nuclear distance \(R = 6.8\) a.u. and time \(t = 24\), and use an initial wave packet localised downfield and upfield respectively. (e) and (f) use a delocalised starting wave packet in the same static field. (e) has \(R = 4\) a.u. and  \(t=29.5\) while (f) has \(R = 14\) a.u. and \(t = 26\). The separatrix of the system is shown by the white line.}
	\label{fig:liouvilleTotal}
\end{figure}

To further expand on this, we now present in Fig.~\ref{fig:liouvilleTotal} the quantum corrections \(Q (x,p,t)\) to the classical Liouville equation [Eq.~(\ref{eq:liouville})]. If the Wigner function has a fully classical time evolution, \(Q (x,p,t)\) vanishes everywhere. As seen in Figure~\ref{fig:liouvilleTotal}, this is not the case. Spots in phase-space that are non-zero indicate where and when the evolution of the Wigner function is non-classical. 

From Fig.~\ref{fig:liouvilleTotal}(a) and (b), we see that this non-classical evolution is not due solely to the electric field, as it is present in the field-free case. By comparing those results to the Wigner quasi-probability density in Fig.~\ref{fig:fieldFree} (c) and (d), the areas around which $Q(x,p,t)= 0.2$  a.u. follow the probability density. This is not true around the central saddle, where the evolution is practically classical. This is expected, as, in this region, the potential barrier may be approximated by an inverted harmonic oscillator. Since such a potential does not contain terms higher than up to the second order, \(Q(x,p,t) = 0\) holds. One should bear in mind, however, that Wigner quasiprobability distributions do show non-local behaviour near separatrices \cite{Balazs1990}.

From adding a static field, we can draw additional conclusions. The quantum corrections are located around the well, close to the region where the quantum bridges occur. If the wave packet is initially placed in the downfield well, the corrections are much weaker, but are still present, as shown in Fig~\ref{fig:liouvilleTotal}(c). This supports the observations in the second column of Fig.~\ref{fig:wignerDownfield}, which shows a faint residual tail towards the upfield centre. 

The more intensely non-classical regions (\(Q(x,p,t) > 0.2 a.u.\)) are located at the quantum bridge, whether it is in the positive or negative momentum region [as seen in Figures~\ref{fig:liouvilleTotal} (d), (e) and (f)]. This also explains why those are absent from Fig~\ref{fig:liouvilleTotal} (c). By starting the initial wave packet downfield and because the inter-nuclear distance is too large, the quantum bridge is very faint.
\begin{figure}[ht]
	\centering
	\includegraphics[width=9cm]{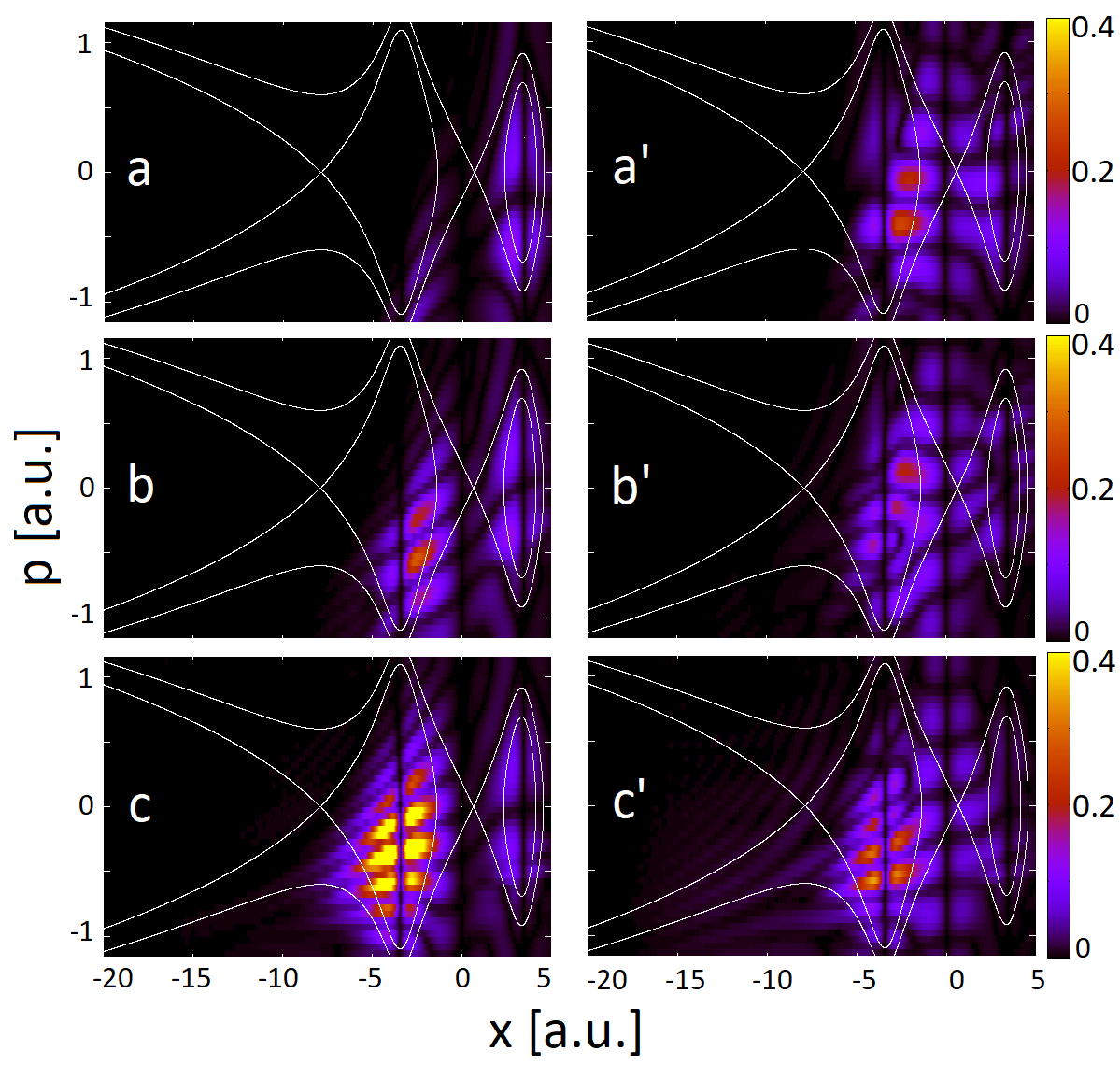}
	\caption{Phase space map of the quantum corrections \(Q(x,p,t)\) (eq.~\ref{eq:liouville}) of a \(H_2^+\) molecule of inter-nuclear distance \(R = 6.8\) in a static laser field of strength \(E=0.0534\) a.u. (intensity \(I=10^{14}  \mathrm{W/cm}^{2})\). The left column uses a localised upfield starting wave packet, while the right column ($'$) a delocalised one. Those are shown at times from top to bottom: \(t = 5\) for (a) and (a$'$), \(t = 10\) for (b) and (b$'$), and \(t = 24\) for (c) and (c$'$). The separatrix of the system is shown by the white line.}
	\label{fig:liouvillePeak}
\end{figure}
Most striking is the very bright %(\( Q(x,p,t) > 0.4 a.u.\)) 
region in Fig~\ref{fig:liouvilleTotal}(d). A comparison with the second column of Fig~\ref{fig:wignerUpfield}(d$'$) shows that it occurs at the quantum bridge that passes through the downfield centre. 

In Fig.~\ref{fig:liouvillePeak}, we compare the evolution of the quantum corrections for an initially delocalised wave packet $\Psi_{\mathrm{cat}}(x,0)$ and a localised upfield wave packet $\Psi_{\mathrm{up}}(x,0)$. There we see that the very high quantum corrections shown in the previous figure build up over time, and are only present when the downfield potential well starts to be populated. The early evolution of both systems, shown in  Figure~\ref{fig:liouvillePeak}(a) and Figure~\ref{fig:liouvillePeak}(a$'$), are radically different. For an initially delocalised wave packet, the quantum bridge as well as the quantum corrections surrounding that bridge are present even at \(t = 5\) a.u.. They fluctuate in time, as seen in Figs.~\ref{fig:liouvillePeak}(b$'$) and (c$'$). On the other hand, for an initial upfield wave packet, the quantum corrections form a steady uphill downhill flow. This supports the argument that there is a quantum mechanism providing a shortcut for the electron to reach the semiclassical escape pathway. 
In both cases, the escape into the continuum appears to be governed by classical dynamics. Indeed the quasi probability flow follows an equienergy curve (as would classical trajectories) far from the core. This is expected as the interaction Hamiltonian is linear in the coordinate $x$ and will be dominant in that region.

\section{Time-dependent fields}
\label{sec:tdep}
\begin{figure}[H]
	\centering
	\includegraphics[width=13cm]{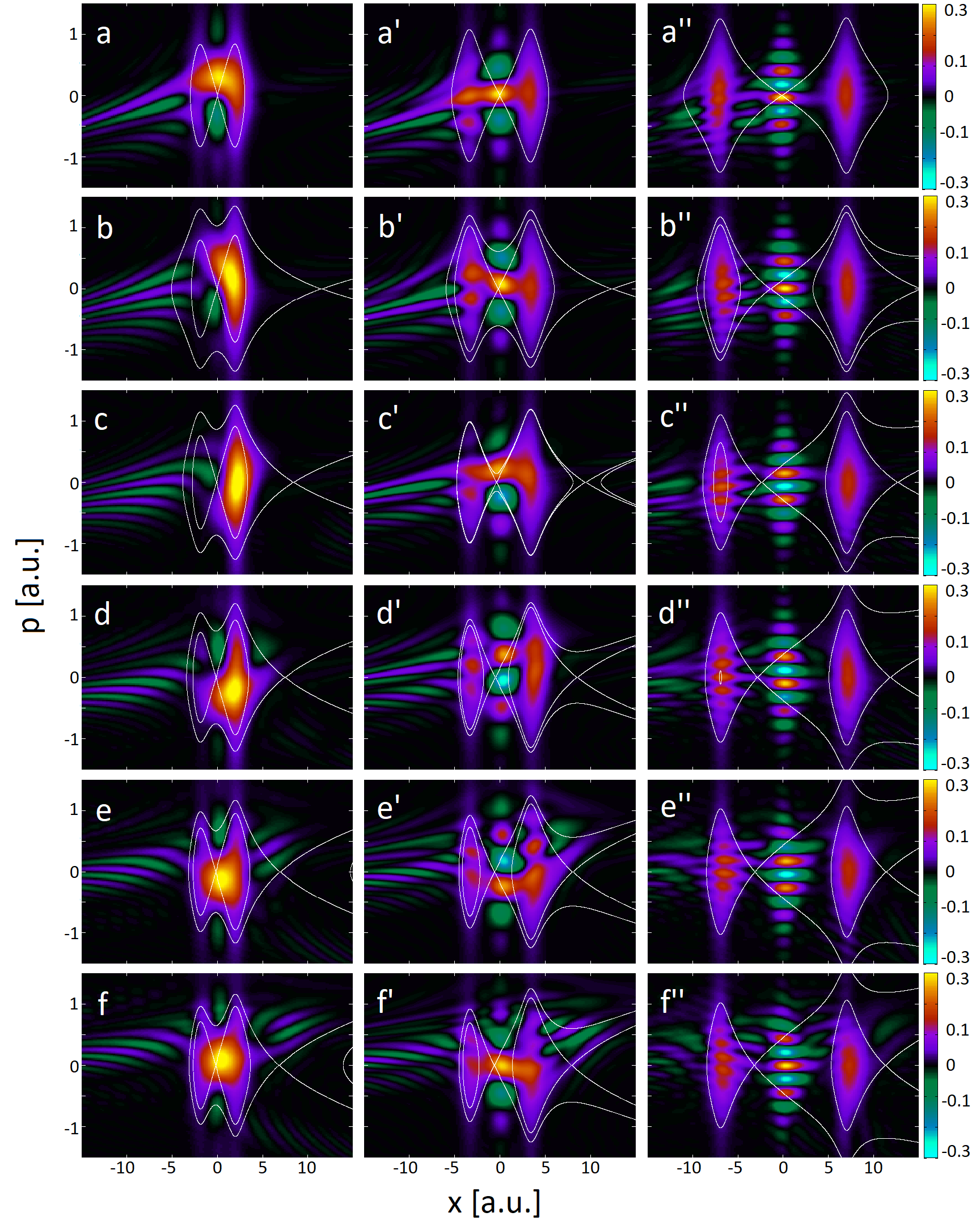}
	\caption{Wigner quasi probability distribution of a model $H_2^+$ molecule in a monochromatic laser field given by Eq.~(\ref{eq:tdfield}) of wavelength \(\lambda= 800 \mathrm{nm}\) and  strength \(E=0.0534\) a.u. (intensity \(I=10^{14}  \mathrm{W/cm}^{2})\) with inter-nuclear distance \(R=4\) (left), \(R=6.8\) (middle) and \(R=14\) (right) at different instants of time: \(t = 0.25T\) [(a), (a$'$) and (a$''$)], \(t = 0.30T\) [(b), (b$'$) and (b$''$)], \(t = 0.35T\) [(c), (c$'$) and (c$''$)],  \(t = 0.40T\) [(d), (d$'$) and (d$''$)],  \(t = 0.45T\)  [(e), (e$'$) and (e$''$)]and \(t = 0.50T\)[(f), (f$'$) and (f$''$)] from top to bottom, where \(T\) is the laser period. }
	\label{fig:wignerTimeDependant}
\end{figure}

After a thorough analysis of ionisation in static and vanishing fields, we can now finally disentangle the effects shown by the Wigner function in a time-dependent field. Thereby, a key issue is how the periodicity of the external field will affect that of the quantum bridges and their cyclic motion. Those are shown in Figure \ref{fig:wignerTimeDependant} using a initial delocalised state and inter-nuclear distances \(R = 4\) a.u., \(R = 6.8\) a.u. and \(R = 14\) a.u.. The Wigner function is plotted over a quarter of the field cycle, from a field crossing [top panels, denoted (a), (a$'$)  and (a$''$)] to a maximum field amplitude [bottom panels, denoted (f), (f$'$)  and (f$''$)] . Prior to that, we have allowed the Wigner function to evolve over a quarter of a cycle, from the previous field extremum to the crossing. This is evidenced by the tails on the left-hand side of the top panels. That way we can emphasise the influence of the changing field gradients on the quasiprobability flow and still carry residual features from the previous evolution.

With the use of a time-dependent field come time-dependent separatrices. At the field crossing [Fig~\ref{fig:wignerTimeDependant}(a)], the separatrix is equal to that of the field-free case. For both \(R = 4\) a.u. and \(R = 14\) a.u. the system is in the same configuration as the static field case, closed and open separatrices respectively. However, for \(R = 6.8\) a.u., the system changes phase-space configurations from nested [Fig~\ref{fig:wignerTimeDependant}(b$'$)] to open [Fig~\ref{fig:wignerTimeDependant}(d$'$)] separatrices in a quarter cycle and spends roughly half of the time in each configuration. 

The shift in momentum gate is again present, and follows the same clockwise cycle discussed in previous sections. The quasiprobability distribution flows via the positive momentum gate from the left to right potential well [see Fig~\ref{fig:wignerTimeDependant}(a) and Fig~\ref{fig:wignerTimeDependant}(c$'$)], and within the same quarter cycle follows the negative momentum gate [see Fig~\ref{fig:wignerTimeDependant}(d) and Fig~\ref{fig:wignerTimeDependant}(e$'$)] from the now downfield (right) to upfield (left) potential wells, seemingly opposing the direction of the field.

There are however key differences, with regard to the static-field case.  Depending on the instantaneous phase-space configuration and on the interplay between the field gradient and the momentum gates, the time-dependence of the field may aid or hinder their clockwise motion and/or ionisation.
For instance, in the first column of Fig.~\ref{fig:wignerTimeDependant}, the bound part of the Wigner function moves to the right [Fig.~\ref{fig:wignerTimeDependant}(a)], but is subsequently hindered by the field gradient to move back towards the left [Fig.~\ref{fig:wignerTimeDependant}(b) and (c)]. This leads to a delay in comparison to the static case displayed in Fig.~\ref{fig:wignerFull}.

\begin{figure}[ht]
	\centering
	\includegraphics[width=10cm]{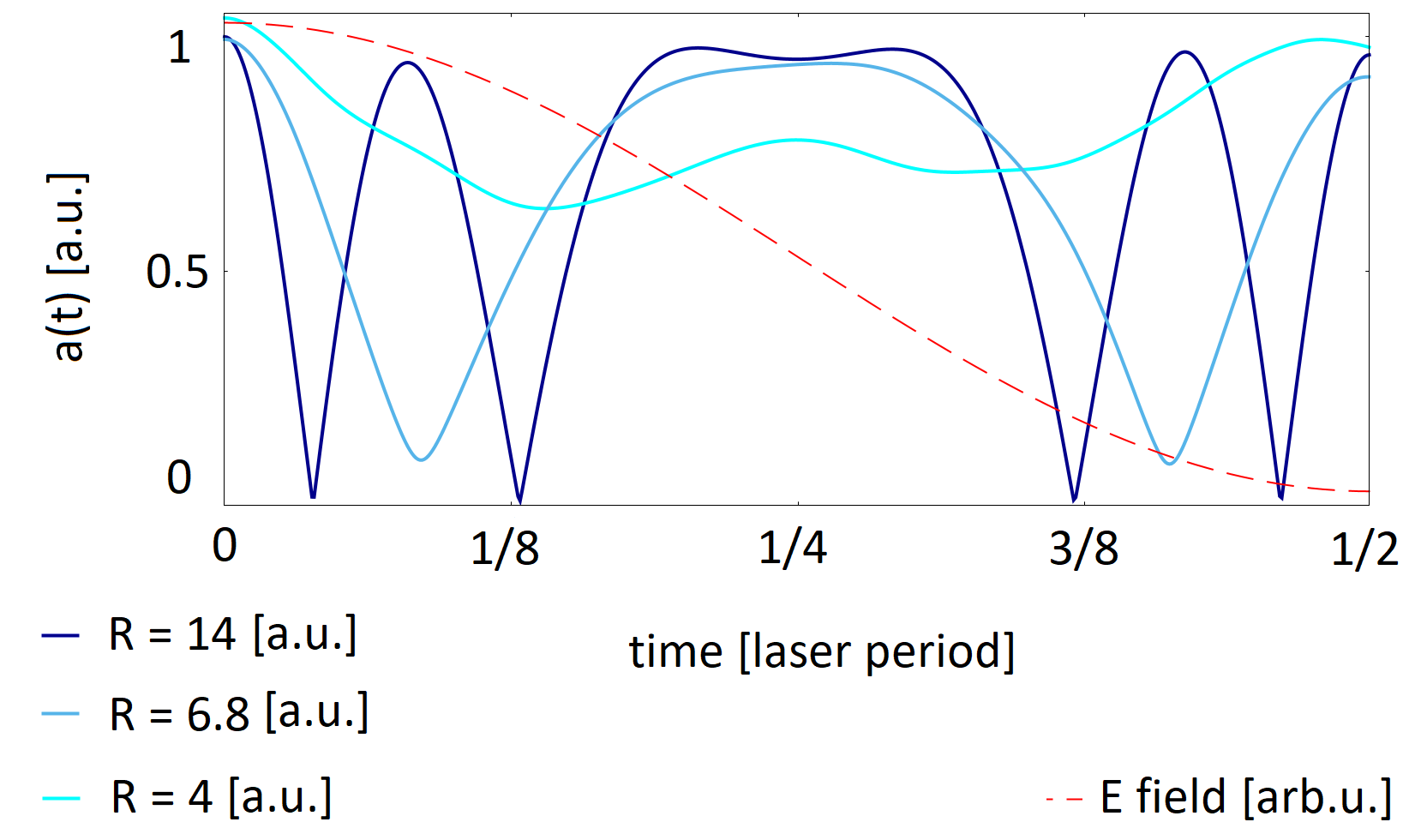}
	\caption{Absolute value of the auto-correlation function over a half-cycle of a monochromatic laser field given by Eq.~(\ref{eq:tdfield}) of wavelength \(\lambda=800 \mathrm{nm}\) and strength \(E = 0.0534\) a.u. (intensity \(I=10^{14}  \mathrm{W/cm}^{2})\), shown in the red dashed line (in arbitrary units) using a delocalised starting wave packet in model $H_2^+$ molecule with inter-nuclear distances of \(R = 4\) a.u. (cyan), \(R = 6.8\) a.u. (dark blue) and \(R = 14\) a.u. (light blue). The time profile of the laser field is indicated by the dashed red line in the figure, but has been shifted and normalised in order to match the scale of the autocorrelation functions.}
	\label{fig:autocorrelationTD}
\end{figure}
If the inter-nuclear distance is such that that the phase-space configuration changes from nested to open separatrices within half a field cycle, the dynamics are more complicated.  A good example is provided in the middle panels of Fig.~\ref{fig:wignerTimeDependant}, for which there is initially a quantum bridge feeding directly into the tails on the left-hand side [Fig.~\ref{fig:wignerTimeDependant}(a$'$)]. When the field changes direction [Fig.~\ref{fig:wignerTimeDependant}(b$'$)], its gradient helps the clockwise motion towards the right. Nonetheless, because the separatrices are nested there will be population trapping, and, consequently an enhancement in the cyclic motion back to the left (upfield) centre, as shown in Fig.~\ref{fig:wignerTimeDependant}(c$'$). The Wigner function will only ``spill" towards the continuum and form tails to the right when the separatrices open [Figs.~\ref{fig:wignerTimeDependant}(d$'$) to (f$'$)]. A remarkable feature in this optimal configuration is that population trapping only occurs when it actually should, i.e., at the times in which the quantum bridges are building up. As the peak-field times are approached, the separatrices open and ionisation bursts occur. However, they do not necessarily follow the field.
For larger inter-nuclear distances (right column in Fig.~~\ref{fig:wignerTimeDependant}), the separatrices are always open but the quantum bridges are rather weak. Therefore ionisation mainly occurs via the quasistatic pathway close to the equienergy curves.

The time evolution of such gates is better studied by looking at the autocorrelation function plotted in Fig.~\ref{fig:autocorrelation} over a half field cycle. Around the maximum and minimum of the laser field, the oscillations of the Wigner function are similar both in frequency and amplitude to those of the static case. Qualitatively, the situation is the same. However, around the field crossing the population has an additional shift, changing direction and following the laser field. It then continues its normal rhythm following the momentum gates around the field minimum. For example, for  \(R = 6.8\) a.u. the population escapes by the negative momentum gate up to around \(T/16\) (where \(T\) is the laser period). The Wigner function then shifts to a positive momentum gate at \(T/8\)(despite the laser field still being in the same direction) going back to its initial distribution. Then, at \(T/4\), where the regular cycle of the Wigner function would create a strong negative momentum gate, the quantum bridge has near zero momentum [Fig~\ref{fig:wignerTimeDependant}(a$'$)] because the direction of the laser field has changed and now counters this movement (instead of adding to it). The autocorrelation function stays more or less constant instead of reaching the minimum associated with a negative momentum gate (see Fig~\ref{fig:autocorrelation} for comparison). The positive momentum gate before \(3T/8\), now following the direction of the laser field, leads the population away from its initial distribution [Fig~\ref{fig:wignerTimeDependant}(c$'$)]. Finally the negative momentum gate (now countering the laser field) shifts the population back to its initial position Fig~\ref{fig:wignerTimeDependant}(e$'$), and the autocorrelation function increases.    

\begin{figure}
    \centering
   \includegraphics[width=14cm]{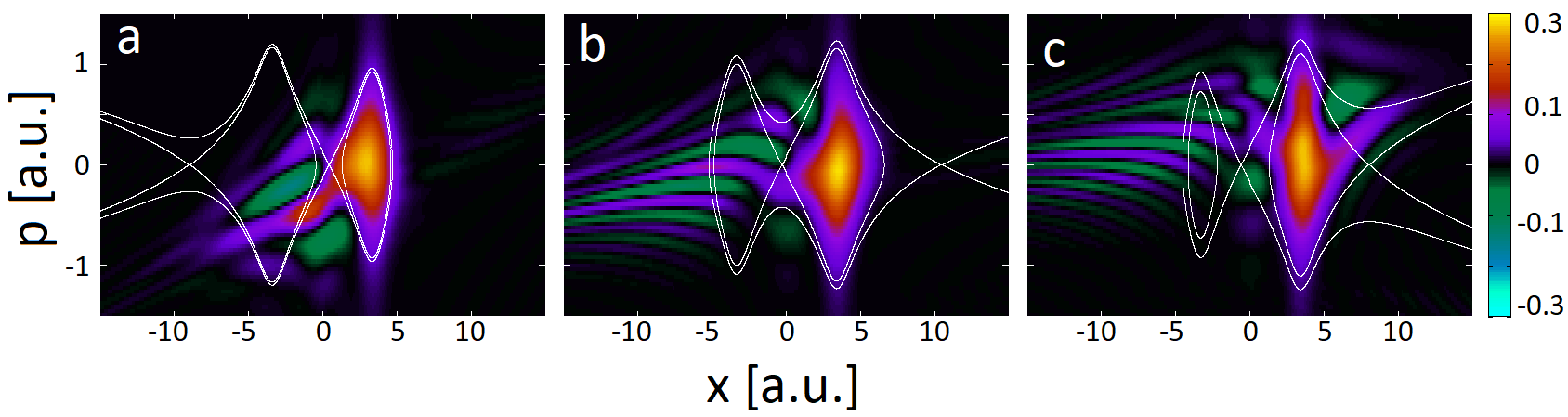}
    \caption{Wigner quasi probability distribution of a model $H_2^+$ molecule in a monochromatic laser field given by Eq.~(\ref{eq:tdfield}) of wavelength \(\lambda= 800 \mathrm{nm}\) and  strength \(E=0.0534\) a.u. (intensity \(I=10^{14}  \mathrm{W/cm}^{2})\) with inter-nuclear distance \(R=6.8\)a.u. and a starting wave packet localised upfield at different instants of time: \(t = 0.13T\) (a), \(t = 0.33T\) (b), \(t = 0.45T\) (c), where \(T\) is the laser period.}
    \label{fig:timedepupfield}
\end{figure}

Finally, if the initial wave packet is localised upfield, there are significant differences from the static case. This is shown in Fig.~\ref{fig:timedepupfield}, for the optimal inter-nuclear distance of $R=6.8$ a.u. During the initial quarter of a cycle,the quantum bridge builds up and directly feeds the ionisation tails [Fig.~\ref{fig:timedepupfield}(a)]. After the field reaches a crossing and switches side, the wave packet will be located in the downhill (right) well. For the time intervals in which the separatrices are nested [Fig.~\ref{fig:timedepupfield}(b)], population trapping will still help a quasiprobability transfer against the field gradient via the momentum gate. However, as field peak is approached and the separatrices are no longer nested [Fig.~\ref{fig:timedepupfield}(c)], this pathway will be strongly suppressed and quasi-static ionisation will prevail (see tails forming on the right-side of the figure). At the subsequent crossing the field will change its direction again, and the quantum bridge towards the left centre will rebuild. Hence, the stationary quantum bridges discussed in Sec.~\ref{sec:Wignerlocalised} are no longer an advantage. Within the timescale of a field cycle, the upfield wave packet does not become a cat state. Longer times will make the interpretation difficult, as, apart from ionisation, there will also be rescattering as an oscillatory field may drive the wave packet back to the core. This will leave marked imprints in the Wigner functions \cite{Kull2012,Baumann2015}.

\section{Conclusions}
\label{sec:conclusions}
In this work, we have performed a detailed analysis of strong-field enhanced ionisation using reduced-dimensionality models of diatomic molecules and phase-space methods, such as Wigner quasiprobability distributions and the quantum Liouville equation. Our studies show that enhanced ionisation stems from the interplay of at least two qualitatively different ionisation pathways, with an optimal phase-space configuration chosen to minimise population trapping and maximise direct downfield population transfer. One of these pathways follows the field gradient and leads to tails along separatrices that ``spill" into the continuum, while the other does not obey field gradients or classical barriers in phase space.  
The former pathway may be associated with quasi-static tunnelling mechanisms \cite{Czirjak2000,Zagoya2014} as well as the semiclassical limit of Wigner quasi-probability distributions \cite{Balazs1990}, with oscillatory tails around separatrices and equienergy curves.  The latter pathway has been first identified in \cite{He2008,Takemo2011} for oscillating driving fields. It consists of a cyclic motion performed by the Wigner function in phase space and the emergence of momentum gates, along which there is a direct quasiprobability flow from one well to the other. Therein, momentum gates were explained as resulting from strongly coupled states and the non-adiabatic response to the time-dependent field gradients. 

We find, however, that this pathway occurs also for static fields, and even in the absence of driving fields altogether. By employing different types of initial bound states for the electronic wave packet, we show that the primary cause of the momentum gates in \cite{He2008,Takemo2011} is quantum interference. If both wells are occupied, quantum interference will create a bridge that will support a direct intra-molecular quasiprobability flow. For initially delocalised (cat) states, quantum bridges are present from the start, while if the electron is initially located in the upfield molecular well they may build up with time. For that, it is necessary that enough quasiprobability density reaches the lower well. This can only happen if the molecular centres are close enough in order to guarantee a significant overlap between the quasi-probability density around the central molecular saddle and that located at the wells. For that very reason, the quantum bridges weaken for increasing inter-nuclear separation. 

The quantum bridges perform a clockwise motion in phase space, whose frequency can be inferred directly using autocorrelation functions, or estimated using classical arguments. The latter requires closed separatrices allowing access to both centres, which, unfortunately, are only present for inter-nuclear separations $R<R_c$, i.e., smaller than that required for enhanced ionisation. Furthermore, the quantum corrections to the Liouville equation are quite large near the quantum bridges. This implies that the classical estimates presented in this work must be viewed with care. Still, they provide the correct direction of evolution for the Wigner functions, and yield reasonable agreement with the empirical quantum values. Away from the molecule, the quantum corrections vanish and the temporal evolution of the Wigner function is essentially classical. 

Depending on the phase-space configuration around the two molecular wells, the quantum bridges may aid or hinder enhanced ionisation. For instance, for  $R<R_c$, there are two nested separatrices and thus significant population trapping. Hence, the downfield population will be forced back to the upfield centre by the quantum bridge's clockwise motion and no enhanced ionisation will occur. In contrast, for larger inter-nuclear distances the outer separatrix will open. This implies that the quantum bridge may strongly connect the population of the upfield centre to the semiclassical escape pathway, thus providing a ``shortcut" that will result in enhanced ionisation. This makes the initial upfield configuration so efficient: in that case, once the quantum bridge has been built, for optimal values of $R$ it may not be able to resume its periodic motion in the uphill direction. This explains why an initial wave packet localised upfield and an inter-nuclear distance of \(R = 6.8\) a.u. leads to the highest ionisation yield: The return is blocked by both the inter-nuclear distance and the lack of population initially in the downfield centre. An initial cat state is less efficient as the strong overlap stimulates the clockwise motion uphill. However, the quantum bridges will still feed the tails that built around various equienergy curves. 

This also sheds light on the behaviour observed for time dependent fields. The frequency of the quantum bridge being higher than that of the laser field, the quasiprobability distribution will sometimes counter-intuitively flow in the direction opposed to the electric-field gradient. The strength of this return is again dependent on the inter-nuclear distance. Remarkably, the frequencies obtained in the present work are within the range of those observed in \cite{Takemo2011}, which is around four times that of a typical near-IR field ($\lambda \simeq 800 \mathrm{nm}$). We have also observed that the quasiprobability flow into the continuum occurs in well-defined temporal bursts, being strongest in the interval for which the separatrices open and there is no population trapping. A key qualitative difference between static and time-dependent fields is that, in the latter case, an initial wave packet localised upfield is no longer preferable to a delocalised (cat) state. This happens because initial upfield states foster the appearance of static quantum bridges, which will be suppressed for a whole quarter of a cycle when the wave packet is upfield. In contrast, initial delocalised states support cyclic bridges building up close to the central saddle, which may be synchronised to the external driving field. 
Finally, the fact that enhanced ionisation is an optimisation problem suggests that the ionisation mechanisms encountered and analysed in this article can be controlled by appropriate coherent superpositions of states, targets and driving fields. This opens up a wide range of possibilities for studying quantum effects in enhanced ionisation. 

\vspace*{0.5cm}

\noindent\textbf{Acknowledgments:} HC acknowledges financial support from the UCL Impact scheme, and CFMF from the UK Engineering and Physical Sciences Research Council (EPSRC; grant no. EP/J019143/1).

\vspace*{0.5cm}

\providecommand{\newblock}{}

\end{document}